\documentclass[aps,twocolumn,pra,showpacs,tightenlines,superscriptaddress,amsmath,amssymb,amsfonts,lengthcheck,articlesuper,article]{revtex4-1}
\usepackage{amsmath}
\usepackage{amsfonts}
\usepackage{graphicx}
\usepackage{epsfig}
\usepackage{color}
\usepackage{txfonts}
\usepackage{bm}
\usepackage[colorlinks,citecolor=blue]{hyperref}
\usepackage{diagbox}
\usepackage{multirow}
\usepackage{array}
\usepackage{makecell}

\begin{document}

\title{Single-photon scattering in giant-atom topological-waveguide-QED systems}
\author{Hai Zhu}
\affiliation{Key Laboratory of Low-Dimensional Quantum Structures and Quantum Control of Ministry of Education, Key Laboratory for Matter Microstructure and Function of Hunan Province, Department of Physics and Synergetic Innovation Center for Quantum Effects and Applications, Hunan Normal University, Changsha 410081, China}
\author{Xian-Li Yin}
\affiliation{Key Laboratory of Low-Dimensional Quantum Structures and Quantum Control of Ministry of Education, Key Laboratory for Matter Microstructure and Function of Hunan Province, Department of Physics and Synergetic Innovation Center for Quantum Effects and Applications, Hunan Normal University, Changsha 410081, China}
\author{Jie-Qiao Liao}
\email{Contact author: jqliao@hunnu.edu.cn}
\affiliation{Key Laboratory of Low-Dimensional Quantum Structures and Quantum Control of Ministry of Education, Key Laboratory for Matter Microstructure and Function of Hunan Province, Department of Physics and Synergetic Innovation Center for Quantum Effects and Applications, Hunan Normal University, Changsha 410081, China}
\affiliation{Institute of Interdisciplinary Studies, Hunan Normal University, Changsha, 410081, China}

\begin{abstract}
The giant-atom topological-waveguide-QED systems have recently emerged as a promising platform for manipulating light-matter interactions. The combination of the multiple-point couplings and topological phase effect could lead to rich physical phenomena and effects. Here, we study single-photon scattering in a Su-Schrieffer-Heeger (SSH) waveguide coupled to either one or two two-level giant atoms. We assume that each giant atom is coupled to the waveguide via two coupling points and hence there exist four and sixteen coupling configurations for the single-giant-atom case and two-giant-atom separate coupling case, respectively. By solving the single-photon scattering problem in the real space, we obtain the exact expressions of the single-photon scattering amplitudes. It is found that a single photon in the SSH waveguide can be completely reflected or transmitted by choosing proper coupling configurations, coupling-point distances, atomic resonance frequency, and dimerization parameter. In addition, under proper parameter conditions, the scattering spectra are periodically modulated by the coupling-point distances. We also find that the collective behavior of the two giant atoms can be adjusted by quantum interference effect and topological effect and that the single-photon scattering spectra can exhibit the Lorentzian, super-Gaussian, electromagnetically induced transparencylike, and asymmetric Fano line shapes for some coupling configurations. This work will inspire the development of controllable single-photon devices based on the giant-atom topological-waveguide-QED systems.
\end{abstract}

\date{\today}
\maketitle
\section{Introduction}
The waveguide quantum electrodynamics (QED) systems~\cite{Sheremet2023,Roy2017,Gu2017}, as an excellent platform for exploring quantum light-matter interactions, have attracted increasing interest due to their potential applications in quantum optics~\cite{Blais1997,Tudela2024} and quantum information science~\cite{Nielsen2000,Zheng2013}. Many interesting physical effects, such as bound state~\cite{John1990,Shen2007,Liao2010a,Liao2010b,Hood2016,Shi2016,Burillo2017,Fong2017,Calajo2019,Mahmoodian2020}, super- and subradiance~\cite{vanLoo2013,Zanner2022}, quantum routers~\cite{Hoi2011,Zhou2013,Shomroni2014}, and few-photon scattering~\cite{Shen2005a,Shen2005b,Chang2007,Zhou2008,Tsoi2008,Liao2009,Shi2009,Tsoi2009,Liao2013,Neumeier2013,Liao2016,Ke2019} have been widely investigated in waveguide-QED systems. In particular, the waveguide-QED systems can serve as good platforms to transfer quantum information and have potential applications in realizing large-scale quantum networks.

In recent years, the giant-atom waveguide-QED systems~\cite{Kockum2014,Kockum125,Kannan2020} have attracted extensive attention from the peers in the fields of quantum optics and quantum information. In most quantum optics systems, atoms are typically treated as pointlike objects~\cite{Walls2008,Scully1997}. This is because the size of natural atoms is much smaller than the wavelengths of the coupled electromagnetic fields. Differently, the size of the giant atoms is comparable to or even larger than the wavelength of the coupled fields. Therefore, the giant atoms can no longer be considered as pointlike objects and then the dipole approximation becomes invalid. Instead, the giant atoms interact with the electromagnetic fields in the waveguide at multiple coupling points and the path quantum interferences become more abundant. So far, many interesting physical phenomena have been found in giant-atom waveguide-QED systems, including frequency-dependent Lamb shift and dissipation rate~\cite{Kockum2014,Vadiraj2021,Yu2021,ZQWang2022,Joshi2023}, decoherence-free interaction between giant atoms~\cite{Kannan2020,Kockum2018,Carollo2020,Soro2023}, non-Markovian dynamics~\cite{Guo2017,Andersson2019,Du2021,Yin2022a,Qiu2023}, single-photon scattering~\cite{Zhao2020,Cai2021,Feng2021,Yin2022b,Zhu2022,Xiang2022,Zhao2024}, and chiral light-matter interactions~\cite{XWang2021,Du2022,Soro2022,XWang2022,Zhou2023,Roccati2024L}. These studies consider the coupling between the giant atoms and either linear or tight-binding waveguides, focusing on the multipoint interference effects. Compared to the linear waveguide, the tight-binding waveguide provides a highly structured environment for the giant atoms and allows them to couple to multiple resonators~\cite{Zhao2020}. In addition, photons propagating in the tight-binding waveguide exhibit a nonlinear dispersion relation and the photonic group velocity can be tuned~\cite{Zhou2008,Zhao2020}.

In parallel, much recent attention in waveguide-QED has been paid to the interaction of atoms with topological waveguides~\cite{Hasan2010,Wra2010,Barik2018,Ozawa2019,Shalaev2019}. Different from the linear and tight-binding waveguides, the SSH waveguide not only provides a highly structured coupling environment for the giant atoms, but also exhibits topological characteristics. Some topics in the small-atom topological-waveguide-QED systems have been studied~\cite{Dusuel2011,Tudela2015,YWang2020,LJWang2020,Kim2021,Dong2021,Leonforte2021,Dong2022,Hauff2022,Vega2023,Roccati2024NC,Lu2024}, such as single-photon scattering~\cite{Barik2018,LJWang2020,Kim2021,Bello2019} and topologically protected quantum entanglement~\cite{Rechtsman2016,Mittal2019,Wang2019,Dai2022}. Meanwhile, people studied the physical phenomena related to the topological effect of waveguide and the interference effect of giant atoms in giant-atom topological-waveguide-QED systems~\cite{Vega2021,Cheng2022,Bag2023,JJWang2023,Luo2024,Wang2024}. Nevertheless, how the topological effect influences the interference effect remains unclear and the various coupling configurations in giant-atom topological waveguide-QED systems are also intriguing. Therefore, when considering the giant atoms coupled to a topological waveguide, the combined effects of both the multipath quantum interference for giant atoms and the topological phase of the waveguide on the single-photon scattering behavior is an interesting topic to be investigated.

In this paper, we study the single-photon scattering in a photonic Su-Schrieffer-Heeger (SSH) waveguide~\cite{Su1979} coupled to either one or two two-level giant atoms. In this case, the incident photon in the SSH waveguide not only has an adjustable group velocity, but also will be influenced by the topological characteristics of the waveguide. By solving the probability amplitude equations in real space, we obtain the exact expressions of the single-photon scattering amplitudes for the two cases. We also find that the scattering amplitudes are determined by the Lamb shift, exchanging interaction, individual decay, and collective decay. The phases of these quantities depend on either the coupling-point distances or the dimerization parameter. It is shown that the single-photon scattering behavior depends on the coupling configurations, coupling-point distances, atomic resonance frequency, and dimerization parameter. In the weak-coupling regime, the scattering spectra can exhibit the Lorentzian, super-Gaussian, electromagnetically induced transparency (EIT)-like~\cite{Abi-Salloum2010,Anisimov2011}, and asymmetric Fano line shapes~\cite{Fano1961,Miroshnichenko2010}. In particular, by adjusting the quantum interference and topological effects, the incident photon can be transmitted completely. In the single-giant-atom case, we analyze in detail the influence of the system parameters on the scattering spectra. We find that the scattering spectra are periodically modulated by the coupling-point distance. In the two-giant-atom case, we discuss the conditions for the appearance of different line shapes and the relationships between the reflection coefficients for different coupling configurations. In particular, we show that when any two or more legs of the giant atoms are coupled to the sublattices of different types, the phases in the scattering amplitudes include the topology-dependent phase. Consequently, the corresponding scattering spectra depend on the topological characteristics of the system.

The rest of this paper is organized as follows. In Secs.~\ref{Single giant atom} and \ref{Two giant atoms}, we study the single-photon scattering in the one- and two-giant-atom topological-waveguide-QED systems, respectively. Concretely, we consider four and sixteen coupling configurations in the single- and two-giant-atom cases, respectively. We solve the stationary Schr\"{o}dinger equation and obtain the scattering amplitudes. We also analyze the dependence of the scattering spectra on the coupling configurations and system parameters. Finally, we conclude this work in Sec.~\ref{Conclusion}.

\section{Single-photon scattering in the single-giant-atom waveguide-QED system}\label{Single giant atom}
In this section, we study single-photon scattering in the single-giant-atom waveguide-QED system. Concretely, we introduce the physical model, present the Hamiltonian, calculate the scattering probability amplitudes, and analyze the scattering spectra.

%%%%%%%%%%%%%%%%%%%%%%%%%%%%%
\begin{figure}[tbp]
\center\includegraphics[width=0.48\textwidth]{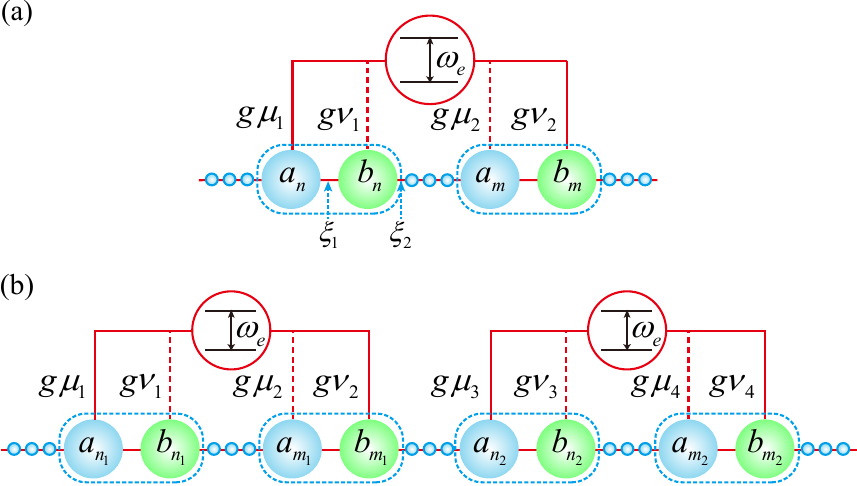}
\caption{(a) Schematic of one two-level giant atom coupled to a Su-Schrieffer-Heeger (SSH) type waveguide via the $n$th and $m$th unit cells. (b) Schematic of two two-level giant atoms separately coupled to an SSH type waveguide via the $n_{1}$th, $m_{1}$th, $n_{2}$th, and $m_{2}$th unit cells.}
\label{model}
\end{figure}
%%%%%%%%%%%%%%%%%%%%%%%%%%%%%
\subsection{Physical model and Hamiltonian}\label{Model and Hamiltonian}
We consider a single two-level giant-atom topological-waveguide-QED system, as shown in Fig.~\ref{model}(a). Here, the giant atom is coupled to a one-dimensional discrete topological waveguide through two separate coupling points and the waveguide is formed by a photonic SSH-type coupled cavity array. The total Hamiltonian of this system reads
\begin{eqnarray}
\hat{H}_{1}=\hat{H}_{a,1}+\hat{H}_{wg}+\hat{H}_{I,1},
\end{eqnarray}
where $\hat{H}_{a,1}$ is the Hamiltonian of the giant atom, $\hat{H}_{wg}$ is the Hamiltonian of the SSH-type waveguide, and $\hat{H}_{I,1}$ describes the interaction between the giant atom and the SSH waveguide.

The Hamiltonian of the giant atom reads ($\hbar=1$)
\begin{eqnarray}
\hat{H}_{a,1}=\omega _{e}\hat{\sigma}^{+}\hat{\sigma}^{-},\label{giant atom}
\end{eqnarray}
where $\omega _{e}$ is the energy separation between the excited state $\left\vert e\right\rangle$ and the ground state $\left\vert g\right\rangle$ of the giant atom. In this work, we set the energy of the ground state $\left\vert g\right\rangle$ as zero. The $\hat{\sigma}^{+}=\left\vert e\right\rangle\left\langle g\right\vert$ and $\hat{\sigma}^{-}=\left\vert g\right\rangle\left\langle e\right\vert$ are, respectively, the raising and lowering operators of the giant atom.

The Hamiltonian of the SSH waveguide is given by
\begin{eqnarray}
\hat{H}_{wg}&=&\omega _{a}\sum_{j=1}^{N}( \hat{a}_{j}^{\dagger }\hat{a}_{j}+%
\hat{b}_{j}^{\dagger }\hat{b}_{j}) -\xi_{1}
\sum_{j=1}^{N}( \hat{a}_{j}^{\dagger }\hat{b}_{j}+\hat{b}_{j}^{\dagger }\hat{a%
}_{j}) \notag \\
&&-\xi_{2} \sum_{j=1}^{N}( \hat{b}_{j}^{\dagger }%
\hat{a}_{j+1}+\hat{a}_{j+1}^{\dagger }\hat{b}_{j}),
\end{eqnarray}
where $\hat{a}_{j}^{\dagger}$ $(\hat{b}_{j}^{\dagger})$ and $\hat{a}_{j}$ $(\hat{b}_{j})$ are, respectively, the creation and annihilation operators of photons at sublattice A (B) in the $j$th ($j = 1,2,3,...,N$) unit cell. The intracell and intercell hopping rates are $\xi_{1}=J\left( 1+\delta \right)$ and $\xi_{2}=J\left( 1-\delta \right)$, respectively, with $\delta$ being the dimerization parameter. For $\delta>0$ ($\delta<0$), the SSH waveguide is in its trivial (topological) phase~\cite{Asboth2016}. To calculate the dispersion relation of the SSH waveguide, we consider the periodic boundary conditions and perform the discrete Fourier transforms
\begin{eqnarray}
\hat{a}_{k}=\frac{1}{\sqrt{N}}\sum_{j=1}^{N}\hat{a}_{j}e^{-ikjd_{0}}, \hspace{0.5cm}
\hat{b}_{k}=\frac{1}{\sqrt{N}}\sum_{j=1}^{N}\hat{b}_{j}e^{-ikjd_{0}}.
\end{eqnarray}
For simplicity, we take the photonic lattice constant $d_{0}=1$. Here, the wave vector in the first Brillouin zone is $k\in[-\pi,...,\pi-2\pi/N]$.

The Hamiltonian of the SSH waveguide can be transformed into $\hat{H}_{wg}=\sum_{k}\hat{Q}_{k}^{\dagger }h_{k}\hat{Q}_{k}$, where we introduce $\hat{Q}_{k}=(\hat{a}_{k},\hat{b}_{k})^{T}$ and the kernel
\begin{eqnarray}
h_{k}=\left(
\begin{array}{cc}
\omega _{a} & y\left( k\right)  \\
y^{\ast }\left( k\right)  & \omega _{a}%
\end{array}%
\right),
\end{eqnarray}
with $y(k)=-(\xi_{1}+\xi_{2}e^{-ik})=\omega_{k}e^{i\phi_{k}}$. Here, $\omega_{k}=J\sqrt{2( 1+\delta^{2}) +2(1-\delta ^{2}) \cos(k)}$ is actually the dispersion relation for the upper energy band, which is insensitive to the sign of $\delta$. Differently, $\phi_{k}=\text{Arg}[-(\xi_{1}+\xi_{2}e^{-ik})]$ is the topology-dependent phase, which is sensitive to the sign of $\delta$. Concretely, the phase $\phi_{k}$ can be divided into $\phi_{k,+}$ and $\phi_{k,-}$, namely,
\begin{eqnarray}
\phi _{k}=\left\{
\begin{array}{cc}
\phi _{k,+}=\text{Arg}[-J(1+\left\vert \delta \right\vert )-J(1-\left\vert
\delta \right\vert )e^{-ik}], & \delta \geq 0, \\
\phi _{k,-}=\text{Arg}[-J(1-\left\vert \delta \right\vert )-J(1+\left\vert
\delta \right\vert )e^{-ik}], & \delta <0,%
\end{array}%
\right.
\end{eqnarray}
which satisfies the condition $\phi_{k,+}+\phi_{k,-}+k=2l\pi$ for an integer $l$. The subscripts ``$+$'' and ``$-$'' represent the sign of $\delta$. For convenience, we set $\omega _{a}$ as the reference of the zero energy. By introducing the eigen-operators
\begin{subequations}
\begin{eqnarray}
\hat{f}_{k}=(\hat{a}_{k}+e^{i\phi_{k}}\hat{b}_{k})/\sqrt{2}, \\
\hat{h}_{k}=(-\hat{a}_{k}+e^{i\phi_{k}}\hat{b}_{k})/\sqrt{2},
\end{eqnarray}
\end{subequations}
the Hamiltonian $\hat{H}_{wg}$ can be diagonalized as
\begin{eqnarray}
\hat{H}_{wg}=\sum_{k}\omega _{k}( \hat{f}_{k}^{\dagger }\hat{f}_{k}-\hat{h}_{k}^{\dagger }\hat{h}_{k}).
\end{eqnarray}
The corresponding dispersion relations for the upper and lower energy bands are, respectively, given by $\omega_{k}$ and $-\omega_{k}$. The middle band gap between the two bands is $4J|\delta|$. Note that there exist two edge states when the SSH waveguide is in its topological phase ($\delta<0$)~\cite{Hasan2010}.

The single giant atom is coupled to the topological waveguide via multiple coupling points. Here, each coupling point is still described by the atom-field interaction under the dipole approximation. We consider the weak-coupling case; then the interaction Hamiltonian can be written under the rotating-wave approximation (RWA) as
\begin{eqnarray}
\hat{H}_{I,1}=g[ ( \mu _{1}\hat{a}_{n}^{\dag }+\nu _{1}\hat{b}%
_{n}^{\dag }+\mu _{2}\hat{a}_{m}^{\dag }+\nu _{2}\hat{b}_{m}^{\dag })
\hat{\sigma}^{-}+\text{H.c.}]. \label{the interaction Hamiltonian}
\end{eqnarray}
In Eq.~(\ref{the interaction Hamiltonian}), we assume that the giant atom is coupled to the SSH waveguide at both the $n$th and $m$th unit cells. For simplicity, we consider that all the coupling points have the same coupling strength $g$. When a giant atom is coupled to an SSH waveguide via two coupling points, there are four different coupling configurations: AA, BB, AB, and BA coupling distributions. To uniformly describe multiple coupling configurations in Fig.~\ref{model}(a), we introduce dimensionless parameters $\mu_{i}$ and $\nu_{i}$ (for $i=1, 2$) in Eq.~(\ref{the interaction Hamiltonian}) to characterize the couplings between the giant atom and the SSH waveguide. The coefficients $\mu_{i}$ and $\nu_{i}$ take $1$ and $0$ corresponding to the presence and absence of the couplings, respectively. Namely, if the giant atom is coupled to the sublattice A (B) at the $j$th cell, we have $\mu _{i}=1$ and $\nu _{i}=0$ ($\mu _{i}=0$ and $\nu _{i}=1$).

\subsection{Scattering probability amplitudes}\label{Scattering probability amplitudes}
In this work, we will use the probability-amplitude method to study the single-photon scattering. In the rotating frame with respect to $\hat{H}_{0}=\omega _{a}\hat{\sigma}^{+}\hat{\sigma}^{-}+\omega _{a}\sum_{j=1}^{N}( \hat{a}_{j}^{\dagger }\hat{a}_{j}+\hat{b}_{j}^{\dagger }\hat{b}_{j})$, the Hamiltonian of the system becomes
\begin{eqnarray}
\hat{H}_{\text{rot},1}&=&\Delta \hat{\sigma}^{+}\hat{\sigma}^{-}-\sum_{j=1}^{N}[\xi_{1}( \hat{a}_{j}^{\dagger }\hat{b}_{j}+\hat{b}_{j}^{\dagger }\hat{a}_{j})+\xi_{2}(\hat{b}_{j}^{\dagger }\hat{a}_{j+1}+\hat{a}_{j+1}^{\dagger }\hat{b}_{j})] \notag \\
&&+g[(\mu _{1}\hat{a}_{n}^{\dag }+\nu _{1}\hat{b}_{n}^{\dag
}+\mu _{2}\hat{a}_{m}^{\dag }+\nu _{2}\hat{b}_{m}^{\dag }) \hat{\sigma}%
^{-}+\text{H.c.}],\label{SSH model}
\end{eqnarray}
where $\Delta=\omega _{e}-\omega _{a}$ is the frequency detuning between the giant atom and the cavity field. The total excitation number operator $\hat{N}=\hat{\sigma}^{+}\hat{\sigma}^{-}+\sum_{j=1}^{N}(\hat{a}_{j}^{\dagger}\hat{a}_{j}+\hat{b}_{j}^{\dagger }\hat{b}_{j})$ of the system is a conserved quantity and thus, in the single-excitation subspace, a general state of the system can be written as
\begin{eqnarray}
\left\vert \Psi _{k}\right\rangle_{1} =c_{e}\left\vert e\right\rangle \left\vert
\emptyset\right\rangle +\sum_{j=1}^{N}[ u_{k}(j) \hat{a}_{j}^{\dagger
}+w_{k}(j) \hat{b}_{j}^{\dagger }] \left\vert
g\right\rangle \left\vert \emptyset\right\rangle, \label{eigenstate1}
\end{eqnarray}
where $\left\vert\emptyset\right\rangle$ indicates that all cavities in the SSH waveguide are in the vacuum state, $c_{e}$ is the probability amplitude corresponding to the single excitation stored in the giant atom, and $u_{k}(j)$ $[w_{k}(j)]$ is the probability amplitude for finding a single photon in the sublattice A (B) of the $j$th cell.

The probability amplitudes can be obtained by solving the stationary Schr\"{o}dinger equation $\hat{H}_{\text{rot},1}\left\vert \Psi _{k}\right\rangle_{1}=\omega^{(\pm)}\left\vert \Psi _{k}\right\rangle_{1}$. To this end, we consider that a single photon with energy $\omega^{(\pm)}=\pm \omega_{k}$ is initially injected from the left-hand side of the waveguide. For the cases of $j<n$, $n<j<m$, and $j>m$, the probability amplitudes $u_{k}(j)$ and $w_{k}(j)$ corresponding to the upper and lower bands take the form~\cite{conditions}
\begin{subequations}
\begin{eqnarray}
u_{k}(j) &=& \left\{
\begin{array}{cc}
e^{ikj}+r_{1}e^{-ikj}, & j<n, \\
Ae^{ikj}+Be^{-ikj}, & n<j<m, \\
t_{1}e^{ikj}, & j>m,%
\end{array}%
\right.\label{probability amplitude u} \\
w_{k}(j) &=& \left\{\,
\begin{array}{cc}
\pm e^{-i\phi _{k}}e^{ikj}\pm e^{i\phi _{k}}r_{1}e^{-ikj}, & j<n, \\
\pm e^{-i\phi _{k}}Ae^{ikj}\pm e^{i\phi _{k}}Be^{-ikj}, & n<j<m, \\
\pm e^{-i\phi _{k}}t_{1}e^{ikj}, & j>m.%
\end{array}%
\right.\label{probability amplitude w}
\end{eqnarray}
\end{subequations}

In Eqs.~(\ref{probability amplitude u}) and (\ref{probability amplitude w}), $r_{1}$ and $t_{1}$ are single-photon reflection and transmission amplitudes, respectively. The variables $A$ and $B$ are the probability amplitudes related to the right- and left-propagating fields between the two coupling points of the giant atom. By using the continuity conditions at the $n$th and $m$th coupling points, we have the relations
\begin{subequations}
\begin{eqnarray}
A&=&1+\frac{\mu _{1}\pm \nu _{1}e^{i\phi _{k}}}{\mu _{1}\pm \nu _{1}e^{-i\phi_{k}}}(r_{1}-B)e^{-2ikn}, \\
t_{1}&=&A+\frac{\mu _{2}\pm \nu _{2}e^{i\phi _{k}}}{\mu _{2}\pm \nu _{2}e^{-i\phi}}Be^{-2ikm}.
\end{eqnarray}
\end{subequations}

The reflection and transmission amplitudes for the single giant-atom case can be obtained as
\begin{subequations}
\begin{eqnarray}
r_{1}^{(\pm)}=\frac{\Gamma^{(\pm)}e^{i\theta^{(\pm )}}e^{2ikn}/2}{i\Delta _{k}^{(\pm)}-i\Delta _{L}^{(\pm)}-\Gamma^{(\pm)}/2}, \label{r} \\
t_{1}^{(\pm)}=\frac{i\Delta _{k}^{(\pm)}-i\Delta _{L}^{(\pm)}}{i\Delta _{k}^{(\pm)}-i\Delta _{L}^{(\pm)}-\Gamma^{(\pm)}/2}. \label{t}
\end{eqnarray}
\end{subequations}

Here, $\Delta _{k}^{(\pm)}=\pm\omega_{k}-\Delta$ are the frequency detunings between the propagating photons in the SSH waveguide (\textquotedblleft$\pm$\textquotedblright\ denoting the upper and lower bands) and the giant atom, $\Delta _{L}^{(\pm )}$ are the Lamb shifts, and $\Gamma^{(\pm )}$ are the effective decay rates. Note that $\theta^{(\pm )}$ are the global phases and do not affect the scattering spectra of the system. These characteristic quantities are defined as
\begin{subequations}
\begin{eqnarray}
\Delta _{L}^{(\pm )} &=&\Gamma_{e}^{(\pm )}\left[(\mu _{1}\mu _{2}+\nu
_{1}\nu _{2})\sin (kd)\mp \sum_{i=1}^{2}\mu _{i}\nu _{i}\sin \phi_{k}\right. \nonumber \\
&&\left. \pm \nu _{1}\mu _{2}\sin (kd+\phi _{k})\pm \mu _{1}\nu _{2}\sin(kd-\phi _{k})\right],  \label{Lamb shift} \\
\Gamma^{(\pm )} &=&2\Gamma_{e}^{(\pm )}\left[\pm \mu _{1}\nu
_{2}\cos (kd-\phi _{k})\pm \nu _{1}\mu _{2}\cos (kd+\phi _{k})\right. \nonumber \\
&&\left. \pm \sum_{i=1}^{2}\mu _{i}\nu _{i}\cos \phi _{k}+(\mu _{1}\mu
_{2}+\nu _{1}\nu _{2})\cos (kd)+1 \right], \nonumber \\
&& \label{individual decay} \\
\theta^{(\pm )} &=&\text{Arg}\left[\Gamma_{e}^{(\pm )}[(\mu _{1}\pm \nu
_{1}e^{-i\phi _{k}})+(\mu _{2}\pm \nu_{2}e^{-i\phi _{k}})e^{ikd}]^{2}\right], \nonumber \\
&& \label{phase}
\end{eqnarray}
\end{subequations}
where $d=m-n$ represents the coupling-point distance and $\Gamma_{e}^{(\pm )}=g^{2}/\upsilon_{g}^{(\pm )}$ are the spontaneous emission rates of the giant atom at each coupling point, with $\upsilon_{g}^{(\pm )} =\partial(\pm\omega _{k})/\partial k=-J^{2}(1-\delta ^{2})\sin k/(\pm\omega _{k})$ being the group velocities of the upper- and lower-band photons in the waveguide.

Equations~(\ref{r}) and (\ref{t}) are the unified expressions of the scattering amplitudes for the AA-, AB-, BA-, and BB-coupling configurations. The reflection and transmission coefficients are defined as $R_{1}^{(\pm)}=|r_{1}^{(\pm)}|^{2}$ and $T_{1}^{(\pm)}=|t_{1}^{(\pm)}|^{2}$, respectively. Due to the conservation of photon number, we have the relation $R_{1}^{(\pm)}+T_{1}^{(\pm)}=1$.

\subsection{Scattering spectra}\label{Scattering spectra}
To study the scattering properties of single photons in the topological waveguide, we investigate the single-photon scattering spectra and analyze the dependence of the scattering spectra on the coupling configurations and system parameters. In particular, we focus on the case where the frequency detuning $\Delta$ lies within the upper band, i.e., $\Delta=\omega _{k}$ with $k\in(-\pi,0)$. Note that the range of $k$ is chosen to ensure that the group velocity of photons in the upper band is positive. Substituting the values of $\mu_{i}$ and $\nu_{i}$ (for $i=1,2$) into Eqs.~(\ref{Lamb shift})-(\ref{phase}), we can obtain the expressions corresponding to the four different coupling configurations. We find that for the AA- and BB-coupling configurations, the two characteristic quantities ($\Gamma^{(\pm )}$ and $\Delta _{L}^{(\pm )}$) have the same form, which results in the same single-photon scattering features. Meanwhile, these quantities are independent of the topology-dependent phase $\phi_{k}$. However, for the AB- and BA-coupling cases, the two quantities depend on both the coupling-point distance $d$ and the phase $\phi_{k}$. Therefore, the single-photon scattering can be jointly adjusted by the quantum interference effect and topological effect.

For convenience, we introduce the accumulated phases $\varphi=\varphi^{\text{AA}}=\varphi^{\text{BB}}=kd$, $\varphi^{\text{AB}}=kd-\phi _{k}$, and $\varphi^{\text{BA}}=kd+\phi _{k}$ of photons propagating between the two coupling points. Here, the superscript represents the different coupling configurations. We find that, for the AA- and BB-coupling cases, the phase is only related to the coupling-point distance $d$, while for the AB- and BA coupling-cases, the phases $\varphi^{\text{AB}}$ and $\varphi^{\text{BA}}$ also include the topology-dependent phase $\phi_{k}$. Therefore, the single-photon scattering behavior for these two coupling configurations is highly sensitive to the topological characteristics of the system. Moreover, we find that the phases $\varphi^{\text{AB}}$ and $\varphi^{\text{BA}}$ satisfy the relation
\begin{eqnarray}
\varphi^{\text{AB}}(d,\delta)-\varphi^{\text{BA}}(d+1,-\delta)=2\pi.
\end{eqnarray}

\begin{figure}[tbp]
\center\includegraphics[width=0.48\textwidth]{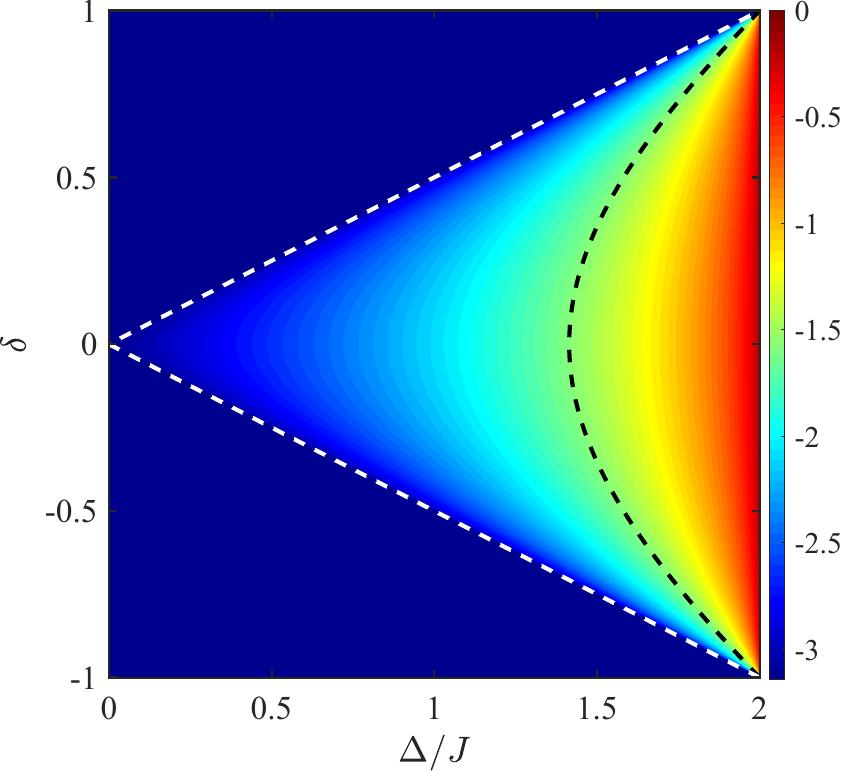}
\caption{Wave vector $k_{\Delta}$ in Eq.~(\ref{wave vector}) as a function of the frequency detuning $\Delta$ and the dimerization parameter $\delta$. The black dashed curve is used to label $k_{\Delta}=-\pi/2$. The area enclosed by the white dashed line determines the range of $\Delta$ and $\delta$.}
\label{kvsDeltaanddelta}
\end{figure}

Based on the above analyses, below we focus on the single-photon scattering in the AA- and AB-coupling cases. Concretely, we analyze the influence of the coupling configurations, coupling-point distances, frequency detuning, and dimerization parameter on the single-photon scattering. Note that the atomic resonance frequency $\Delta$ can lie within the band gap, inside the energy bands, or outside the energy bands. Some previous studies of the giant atom-topological waveguide QED systems have shown that atom-photon bound states can appear in the system when the atomic resonance frequency lies within the band gap or outside the bands~\cite{Bello2019,Vega2021,Cheng2022,Wang2024}. In particular, when the atomic resonance frequency lies within the band gap (for example $\Delta=0$), the giant atom can act as an effective boundary and induce the chiral zero-energy modes (i.e., the chiral bound states) in the waveguide under the periodic boundary condition~\cite{Cheng2022}. The properties of these chiral bound states are similar to the edge states under the open-boundary SSH waveguide, but their formation conditions and distributions are different~\cite{Bello2019,Vega2021,Cheng2022,Wang2024}. First, the chiral bound states and the edge states correspond to  periodic and open boundary conditions of the waveguide, respectively. Second, the photonic probability distributions of chiral bound states are located at the left side of the giant atom or between the two coupling points for A-A coupling~\cite{Cheng2022}. The edge states are mostly located at the two ends of the SSH waveguide~\cite{Cheng2022,Lang2012}. Here, we focus on the single-photon scattering with the incident-photon frequency within the energy band. It is found that when the atomic resonance frequency lies within the band gap or outside the energy band, the incident photon is completely transmitted. In contrast, when the atomic resonance frequency is within the energy band and resonates with the incident photon ($\omega _{k}=\Delta$), the photon can be scattered. In this case, the wave vector $k$ can be expressed as
\begin{eqnarray}
k_{\Delta}=-\left\vert\arccos\left[\frac{\Delta ^{2}-2J^{2}(
1+\delta ^{2}) }{2J^{2}( 1-\delta ^{2}) }\right]
\right\vert. \label{wave vector}
\end{eqnarray}
According to Eq.~(\ref{wave vector}), the group velocity $\upsilon_{g}$ is denoted as $\upsilon_{g,k_{\Delta }}$. The spontaneous emission rate becomes $\Gamma _{e}(\Delta,\delta)=g^{2}/\upsilon_{g,k_{\Delta }}$. To clearly see the dependence of $k_{\Delta}$ on the detuning $\Delta$ and the dimerization parameter $\delta$, in Fig.~\ref{kvsDeltaanddelta} we plot the wave vector $k_{\Delta}$ as a function of $\Delta$ and $\delta$. We can see that the range of the wave vector is $k_{\Delta}\in(-\pi,0)$. For the resonant case $\omega _{k}=\Delta$, the values of $\Delta$ and $\delta$ satisfying Eq.~(\ref{wave vector}) fall within the triangular region delineated by the white dashed lines in Fig.~\ref{kvsDeltaanddelta}.
%%%%%%%%%%%%%%%%%%%%%%%%%%%%%
%%%%%%%%%%%%%%%%%%%%%%%%%%%%%

\begin{figure}[tbp]
\center\includegraphics[width=0.46\textwidth]{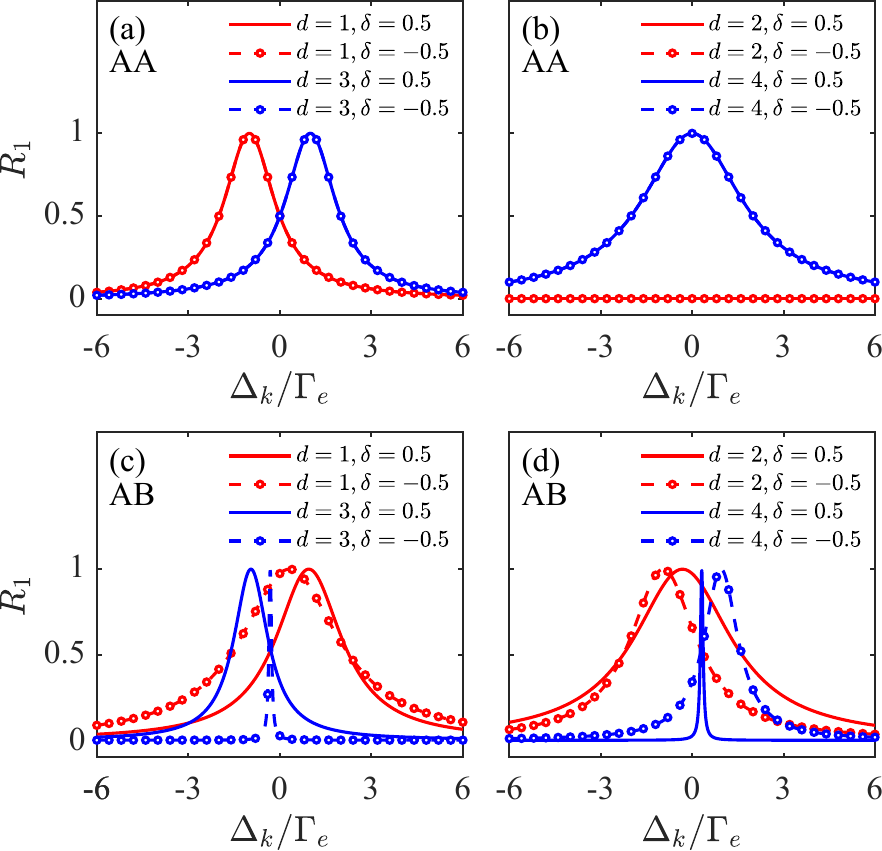}
\caption{Reflection coefficient $R_{1}$ as a function of the detuning $\Delta_{k}$ at the trivial phase ($\delta>0$) and the topological phase ($\delta<0$) for (a) and (b) the AA-coupling configuration and (c) and (d) the AB-coupling configuration. The left and right columns correspond to the odd and even coupling-point distances, respectively. Other parameters used are $g=0.01J$ and $\Delta\approx1.58J$ ($k_{\Delta}=-\pi/2$).}
\label{RAAandABvsDeltak}
\end{figure}

We first study the influence of the sign of the dimerization parameter $\delta$ and the coupling-point distance $d$ on the single-photon scattering. Specifically, we take $\delta=\pm0.5$ and $k_{\Delta}=-\pi/2$. According to Eq.~(\ref{wave vector}), the corresponding frequency detuning is $\Delta\approx1.58J$. In Fig.~\ref{RAAandABvsDeltak}, we plot the reflection coefficient $R_{1}$ for the AA- and AB-coupling configurations as a function of the detuning $\Delta_{k}$ when $\delta$ and $d$ take different values. For the AA-coupling configuration, the sign of $\delta$ has no influence on the reflection spectra, which is consistent with the previous analyses. However, the locations of the reflection peaks can be adjusted by changing the values of $d$. In particular, when $d=1$ and $3$, we find that the reflection spectra are the Lorentzian line shapes, as shown in Fig.~\ref{RAAandABvsDeltak}(a). We also find that the two Lorentzian line shapes are symmetric with respect to $\Delta_{k}=0$ and have equal linewidths. When $d=2$ and $4$, the reflection spectra in Fig.~\ref{RAAandABvsDeltak}(b) exhibit remarkably different features. For $d=4$, the reflection spectrum exhibits the Lorentzian line shape with reflection peak located at $\Delta_{k}=0$. Differently, for $d=2$, the incident photon is completely transmitted ($R_{1}=0$). This phenomenon is caused by the constructive interference between the two coupling channels of the giant atom~\cite{Kockum125}.
%%%%%%%%%%%%%%%%%%%%%%%%%%%%%
%%%%%%%%%%%%%%%%%%%%%%%%%%%%%

For the AB-coupling configuration, we find that the reflection spectra are characterized by the Lorentzian line shapes, as shown in Figs.~\ref{RAAandABvsDeltak}(c) and \ref{RAAandABvsDeltak}(d). Different from the AA-coupling configuration, both the sign of $\delta$ and the value of $d$ can affect the reflection spectra of the AB-coupling configuration. Though the reflection spectra are the Lorentzian line shapes for both $\delta=0.5$ and $-0.5$, the sign of $\delta$ can change the reflection peak and linewidth. This means that the single-photon scattering for the AB-coupling configuration is topologically dependent. Moreover, we find that the red solid curve in Fig.~\ref{RAAandABvsDeltak}(c) (with $\delta=0.5$ and $d=1$) and the red dashed curve with dots in Fig.~\ref{RAAandABvsDeltak}(d) (with $\delta=-0.5$ and $d=2$) are symmetric with respect to $\Delta_{k}=0$ and vice versa. These two reflection spectra have equal linewidths. The same relationship is observed for the reflection spectra for $d=3$ and $d=4$ [see the blue solid curve and dashed curve with dots in Figs.~\ref{RAAandABvsDeltak}(c) and \ref{RAAandABvsDeltak}(d)]. We should mention that the dependence of the single-photon scattering on the topological feature of the waveguide can also be studied by examining the dependence of the scattering behavior on the Zak phase. A relating experimental study has been realized in periodic acoustic systems~\cite{Xiao2015}. The relation between the scattering feature and the Zak phase can be directly obtained based on the correspondence between $\delta>0$ ($\delta<0$) and the $0$ ($\pi$) Zak phase.
%%%%%%%%%%%%%%%%%%%%%%%%%%%
\begin{table*}[ht]
    \renewcommand{\arraystretch}{0.95}
    \caption{The Lamb shifts and decay rates of the single giant atom coupled to the SSH waveguide in the AA- and AB-coupling configurations.}
    \label{table_two}
    \begin{ruledtabular}
        \begin{tabular*}{\linewidth}{ccccccccc}
        \makecell[l]{Coupling\\ configurations} & \makecell[l]{Lamb shifts and\\ decay rates} & \diagbox[height=2.2em,width=2.5em]{$\delta$}{$d$} & $4x$ & $4x+1$ & $4x+2$ & $4x+3$ & $4x+4$ \\
        \hline
        AA & $\Delta _{L,k_{\Delta}}^{\text{AA}}$ & $\pm0.5$ & & $-\Gamma _{e}$ & $0$ & $\Gamma _{e}$ & $0$ \\
         & $\Gamma_{k_{\Delta}}^{\text{AA}}$ & $\pm0.5$ & & $2\Gamma _{e}$ & $0$ & $2\Gamma _{e}$ & $4\Gamma _{e}$ \\
        AB & $\Delta _{L,k_{\Delta}}^{\text{AB}}$ & 0.5 & $-\Gamma _{e}\sin \phi _{k_{\Delta},+}$ & $-\Gamma _{e}\cos \phi _{k_{\Delta},+}$ & $\Gamma _{e}\sin \phi _{k_{\Delta},+}$ & $\Gamma _{e}\cos \phi _{k_{\Delta},+}$ & $-\Gamma _{e}\sin \phi _{k_{\Delta},+}$ \\
         & & $-0.5$ & $-\Gamma _{e}\cos \phi _{k_{\Delta},+}$ & $-\Gamma _{e}\sin \phi _{k_{\Delta},+}$ & $\Gamma _{e}\cos \phi _{k_{\Delta},+}$ & $\Gamma _{e}\sin \phi _{k_{\Delta},+}$ & $-\Gamma _{e}\cos \phi _{k_{\Delta},+}$ \\
        & $\Gamma_{k_{\Delta}}^{\text{AB}}$ & 0.5 &$2\Gamma _{e}(1+\cos \phi _{k_{\Delta},+})$ & $2\Gamma _{e}(1-\sin \phi _{k_{\Delta},+})$ & $2\Gamma _{e}(1-\cos \phi _{k_{\Delta},+})$ & $2\Gamma _{e}(1+\sin \phi _{k_{\Delta},+})$ & $2\Gamma _{e}(1+\cos \phi _{k_{\Delta},+})$ \\
        & & $-0.5$ & $2\Gamma _{e}(1+\sin \phi _{k_{\Delta},+})$ & $2\Gamma _{e}(1-\cos \phi _{k_{\Delta},+})$ & $2\Gamma _{e}(1-\sin \phi _{k_{\Delta},+})$ & $2\Gamma _{e}(1+\cos \phi _{k_{\Delta},+})$ & $2\Gamma _{e}(1+\sin \phi _{k_{\Delta},+})$ \\
        \end{tabular*}
    \end{ruledtabular}
    \label{table:Lamb shift and individual decay}
\end{table*}
%%%%%%%%%%%%%%%%%%%%%%%%%%%%%%%%

To better explain these phenomena, we only consider the case where the incident photons in the vicinity of atomic transition frequency can effectively interact with the atom. Consequently, we can replace the wave vector $k$ in Eqs.~(\ref{Lamb shift})--(\ref{phase}) with $k_{\Delta}$. In this case, the characteristic quantities in Eqs.~(\ref{Lamb shift})--(\ref{phase}) for the upper-band case can be approximated as
\begin{subequations}
\begin{eqnarray}
\Delta _{L}^{(+)} &\approx& \Delta _{L,k_{\Delta}} =\Gamma _{e}\left[( \mu _{1}\mu _{2}+\nu _{1}\nu _{2}) \sin\varphi_{k_{\Delta}} +\mu _{1}\nu _{2}\sin\varphi_{k_{\Delta}}^{\text{AB}}\right. \nonumber \\
&&\left.+\nu _{1}\mu _{2}\sin\varphi_{k_{\Delta}}^{\text{BA}}-\sum_{i=1}^{2}\mu _{i}\nu _{i}\sin \phi_{k_{\Delta}}\right], \label{approximated:Lamb shift} \\
\Gamma^{(+)} &\approx& \Gamma_{k_{\Delta}} =2\Gamma _{e}\left[( \mu _{1}\mu _{2}+\nu _{1}\nu _{2}) \cos\varphi_{k_{\Delta}} +\mu _{1}\nu _{2}\cos\varphi_{k_{\Delta}}^{\text{AB}}\right. \nonumber \\
&&\left.+\nu _{1}\mu _{2}\cos\varphi_{k_{\Delta}}^{\text{BA}}+\sum_{i=1}^{2}\mu _{i}\nu _{i}\cos \phi_{k_{\Delta}}+ 1 \right], \label{approximated:individual decay} \\
\theta_{k}^{(+)} &\approx & \theta_{k_{\Delta}} =\text{Arg}\left[\Gamma _{e}[(\mu _{1}+ \nu
_{1}e^{-i\phi _{k_{\Delta}}})+(\mu _{2}+ \nu_{2}e^{-i\phi _{k_{\Delta}}})e^{ik_{\Delta}d}]^{2}\right], \nonumber \\
&& \label{approximated:phase}
\end{eqnarray}
\end{subequations}
where the corresponding propagating phases are $\varphi_{k_{\Delta}}=k_{\Delta}d$, $\varphi_{k_{\Delta}}^{\text{AB}}=k_{\Delta}d-\phi_{k_{\Delta}}$, and $\varphi_{k_{\Delta}}^{\text{BA}}=k_{\Delta}d+\phi_{k_{\Delta}}$. According to Eqs.~(\ref{approximated:Lamb shift})--(\ref{approximated:phase}), Eq.~(\ref{r}) can be approximated as
\begin{eqnarray}
R_{1}=R_{1}^{(+)}=|r_{1}^{(+)}|^{2}\approx\frac{\Gamma_{k_{\Delta}}^{2}/4}{(\Delta _{k}-\Delta _{L,k_{\Delta}})^{2}+\Gamma_{k_{\Delta}}^{2}/4}. \label{approximated:r}
\end{eqnarray}
Under the same parameter conditions, the approximated reflection spectra coincide with the original reflection spectra.

By substituting $k_{\Delta}=-\pi/2$ (with $\Delta\approx1.58J$) into Eqs.~(\ref{approximated:Lamb shift})--(\ref{approximated:phase}), we can further obtain the Lamb shifts and decay rates of the giant atom for the AA- and AB-coupling configurations under different values of $\delta$ and $d$ and these characteristic quantities can be used to explain the features of the reflection spectra. The concrete expressions are summarized in Table~\ref{table:Lamb shift and individual decay}. It can be seen that the Lamb shifts and decay rates are periodically modulated by the coupling-point distance, with a period of $d=4$. When we take $d=4x+x_{0}$ (with $x_{0}=1$--$4$ for the AA-coupling case), the reflection coefficients in Eq.~(\ref{approximated:r}) become
\begin{eqnarray}
\label{CS-ss-eg-ge}
R_{1}^{\text{AA}}\approx & \left\{ \begin{array}{c}
\Gamma _{e}^{2}/[(\Delta _{k}+\Gamma _{e})^{2}+\Gamma _{e}^{2}],\\[1.0ex]
0,\\[1.2ex]
\Gamma _{e}^{2}/[(\Delta _{k}-\Gamma _{e})^{2}+\Gamma _{e}^{2}],\\[1.4ex]
(2\Gamma _{e})^{2}/[(\Delta _{k})^{2}+(2\Gamma _{e})^{2}],
\end{array}\right. & \hspace{0.3cm}\begin{split}
d & =4x+1,\\[0.5ex]
d & =4x+2,\\[0.5ex]
d & =4x+3,\\[0.5ex]
d & =4x+4, \label{reflection coefficients of AA}
\end{split}
\end{eqnarray}
where the superscript represents the AA-coupling case. Equation~(\ref{reflection coefficients of AA}) indicates that, when $d=4x+1$, $4x+3$, and $4x+4$, the reflection spectra are characterized by the Lorentzian line shapes centered at $\Delta _{k}=-\Gamma _{e}$, $\Gamma _{e}$, and $0$ with the linewidth of $2\Gamma _{e}$, $2\Gamma _{e}$, and $4\Gamma _{e}$, respectively. These results indicate that the approximation is valid. For $d=4x+2$, we have $\Delta _{L,k_{\Delta}}^{\text{AA}}=0$ and $\Gamma_{k_{\Delta}}^{\text{AA}}=0$. In this case, the giant atom is decoupled from the SSH waveguide, resulting in a complete transmission of the incident photon in the waveguide [the red solid curve and dashed curve with dots in Fig.~\ref{RAAandABvsDeltak}(b)].

For the AB-coupling configuration, both the Lamb shifts and decay rates in Table~\ref{table:Lamb shift and individual decay} are nonzero for all $d=4x+x_{0}$ (with $x_{0}=0$--$3$) when $|\delta|=0.5$ and $k _{\Delta}=-\pi/2$. This makes the reflection spectra exhibit the Lorentzian line shapes centered at $\Delta _{k}=\Delta _{L,k_{\Delta}}^{\text{AB}}$ with spectrum linewidth of $\Gamma_{k_{\Delta}}^{\text{AB}}$, which is consistent with numerical results in Figs.~\ref{RAAandABvsDeltak}(c) and \ref{RAAandABvsDeltak}(d). Specifically, when $d=4x+1$ and $\delta=0.5$ [the red solid curve in Fig.~\ref{RAAandABvsDeltak}(c)], the Lamb shift and decay rate are given by $-\Gamma _{e}\cos \phi _{k_{\Delta},+}$ and $2\Gamma _{e}(1-\sin \phi _{k_{\Delta},+})$, respectively. When $d=4x+2$ and $\delta=-0.5$ [the red dashed curve with dots in Fig.~\ref{RAAandABvsDeltak}(d)], the Lamb shift and decay rate are, respectively, given by $\Gamma _{e}\cos \phi _{k_{\Delta},+}$ and $2\Gamma _{e}(1-\sin \phi _{k_{\Delta},+})$. We find that the two reflection spectra have opposite Lamb shifts and the same decay rates. For $d=4x+3$ (with $\delta=0.5$) and $d=4x+4$ (with $\delta=-0.5$), the characteristic quantities of the reflection spectra exhibit opposite Lamb shifts and the same decay rates, as shown by the blue solid curve in Fig.~\ref{RAAandABvsDeltak}(c) and the blue dashed curve with dots in Fig.~\ref{RAAandABvsDeltak}(d).

\begin{figure}[tbp]
\center\includegraphics[width=0.48\textwidth]{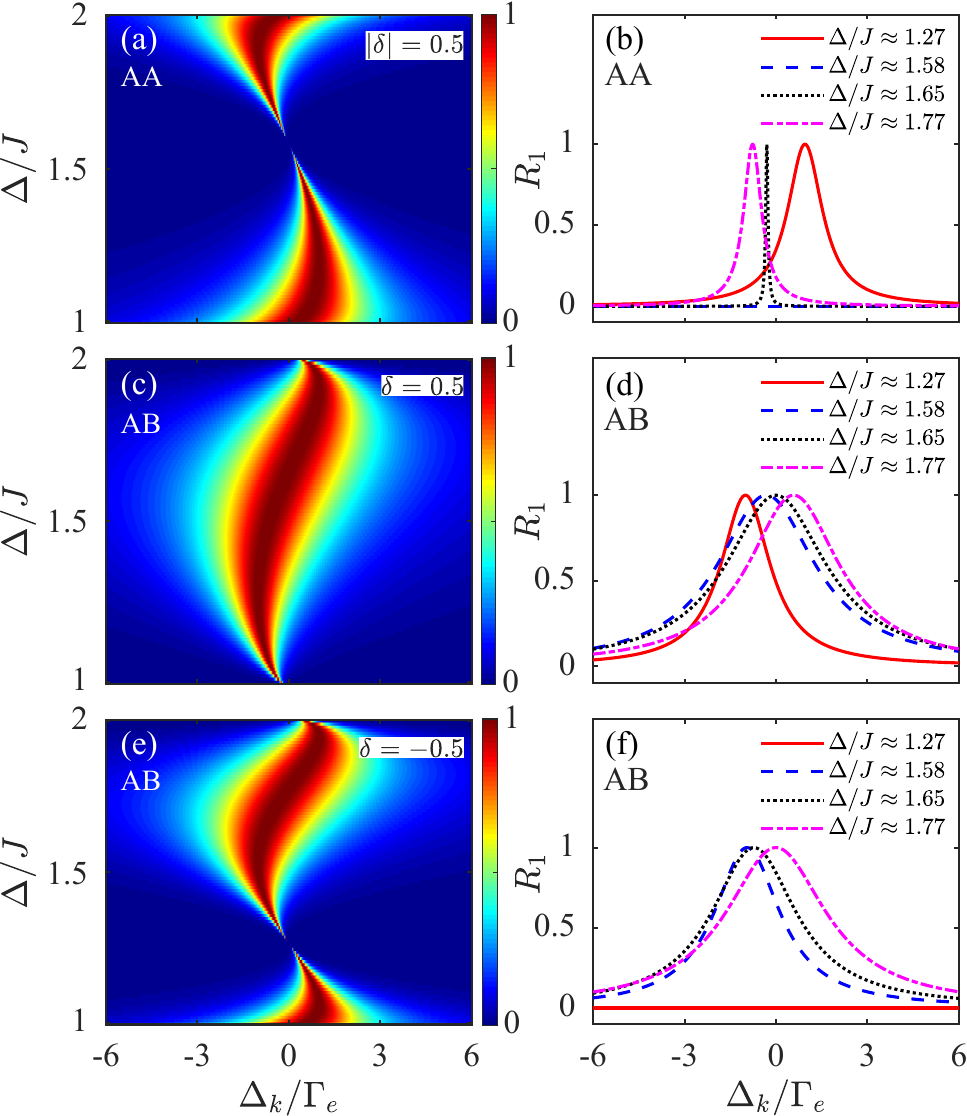}
\caption{Reflection coefficient $R_{1}$ as a function of $\Delta _{k}$ and $\Delta$ for different coupling configurations and various values of $\delta$: (a) AA coupling and $|\delta|=0.5$, (c) AB coupling and $\delta=0.5$, and (e) AB coupling and $\delta=-0.5$. The profiles of panels (a), (c), and (e) at several values of $\Delta$ are shown by the curves in (b), (d), and (f). Here, the red solid, blue dashed, black dotted, and purple dot-dashed curves correspond to $\Delta\approx1.27J$, $\Delta\approx1.58J$, $\Delta\approx1.65J$, and $\Delta\approx1.77J$, respectively. Other parameters used are $g=0.01J$ and $d=2$.}
\label{RAAandABvsDeltakchgDelta}
\end{figure}

To see the influence of the atom transition frequency and the dimerization parameter on the single-photon scattering, in Figs.~\ref{RAAandABvsDeltakchgDelta}(a), \ref{RAAandABvsDeltakchgDelta}(c), and  \ref{RAAandABvsDeltakchgDelta}(e) we show the reflection coefficient as a function of $\Delta$ and $\Delta _{k}$ when $d$ takes a fixed value. For more details, we also plot in Figs.~\ref{RAAandABvsDeltakchgDelta}(b), \ref{RAAandABvsDeltakchgDelta}(d), and \ref{RAAandABvsDeltakchgDelta}(f) the profiles at some typical values of the detuning in the region of $\Delta\in(J,2J)$. We point out that the reflection spectra of AA-coupling configuration is independent of the sign of $\delta$. Thus we only need to analyze the influence of $|\delta|$ on the scattering spectra. From Figs.~\ref{RAAandABvsDeltakchgDelta}(a) and \ref{RAAandABvsDeltakchgDelta}(b) (AA-coupling case), the detuning $\Delta$ plays an important role in the single-photon scattering process. The linewidth $\Gamma_{k_{\Delta}}^{\text{AA}}$ of the reflection spectrum gradually narrows as $\Delta$ increases from $J$ to $1.58J$ and then progressively widens as $\Delta$ increases from $1.58J$ to $2J$. In particular, the reflection peak disappears completely when $\Delta\approx1.58J$, as shown by the blue dashed curve in Fig.~\ref{RAAandABvsDeltakchgDelta}(b). This is because the giant atom is decoupled from the SSH waveguide. In addition, the scattering behavior is independent of the sign of $\delta$, which confirms our previous analyses.

%%%%%%%%%%%%%%%%%%%%%%%%%%%%%
%%%%%%%%%%%%%%%%%%%%%%%%%%%%
Different from the AA-coupling case, the single-photon scattering behavior for the AB-coupling case is not only modulated by the detuning $\Delta$, but also depends on the sign of $\delta$. As shown in Figs.~\ref{RAAandABvsDeltakchgDelta}(c), \ref{RAAandABvsDeltakchgDelta}(d), \ref{RAAandABvsDeltakchgDelta}(e), and \ref{RAAandABvsDeltakchgDelta}(f), the reflection spectra at $\delta=0.5$ and $\delta=-0.5$ indicate different dependencies on $\Delta$. In Figs.~\ref{RAAandABvsDeltakchgDelta}(c) and \ref{RAAandABvsDeltakchgDelta}(d), when $\delta=0.5$, the linewidth $\Gamma_{k_{\Delta}}^{\text{AB}}$ of the reflection spectrum becomes wide as $\Delta$ increases from $J$ to $1.65J$ and then becomes narrow as $\Delta$ increases from $1.65J$ to $2J$. When $\Delta\approx1.65J$, the linewidth of the reflection spectrum reaches its maximum value, as shown by the black dotted curve in Fig.~\ref{RAAandABvsDeltakchgDelta}(d). Meanwhile, the reflection peak is located at $\Delta _{k}=0$. For $\delta=-0.5$, as shown in Figs.~\ref{RAAandABvsDeltakchgDelta}(e) and \ref{RAAandABvsDeltakchgDelta}(f), the dependence of the linewidths for the reflection spectra on the detuning $\Delta$ can be divided into three regions: (i) $\Gamma_{k_{\Delta}}^{\text{AB}}$ decreases monotonically with $\Delta$ in the region $\Delta\in(J,1.27J)$, (ii) $\Gamma_{k_{\Delta}}^{\text{AB}}$ increases monotonically with $\Delta$ in the region $\Delta\in[1.27J,1.77J)$ and (iii) $\Gamma_{k_{\Delta}}^{\text{AB}}$ decreases monotonically with $\Delta$ in the region $\Delta\in[1.77J,2J)$. In particular, the reflection peak disappears completely when $\Delta\approx1.27J$, as shown by the red solid curve in Fig.~\ref{RAAandABvsDeltakchgDelta}(f). This is also caused by the decoupling between the giant atom and the waveguide. In addition, the width of the reflection spectrum is the widest when $\Delta\approx1.77J$, as shown by the purple dot-dashed curve in Fig.~\ref{RAAandABvsDeltakchgDelta}(f).

\section{Single-photon scattering in the two-giant-atom waveguide-QED system}\label{Two giant atoms}
In this section, we study single-photon scattering in the two-giant-atom waveguide-QED system. Concretely, we introduce the physical model and the Hamiltonian, derive the scattering probability amplitudes, and analyze the scattering spectra.

\subsection{Physical model and Hamiltonian}
The two-giant-atom waveguide-QED system under consideration consists of two giant atoms coupled to an SSH-type topological waveguide. In this work we only consider the separate-coupling case for the two giant atoms. In the rotating frame with respect to $\hat{H}_{0}^{'}=\omega_{a}\sum_{s=1}^{2}\hat{\sigma}_{s}^{+}\hat{\sigma}_{s}^{-}+\omega _{a}\sum_{j=1}^{N}( \hat{a}_{j}^{\dagger }\hat{a}_{j}+\hat{b}_{j}^{\dagger }\hat{b}_{j})$, the total Hamiltonian of the system reads~\cite{Luo2024}
\begin{eqnarray}
\hat{H}_{\text{rot},2}=\hat{H}_{a,2}+\hat{H}_{wg}+\hat{H}_{I,2},\label{H}
\end{eqnarray}
where the Hamiltonian of the SSH waveguide has been introduced by the second term of Eq.~(\ref{SSH model}). We consider two degenerate giant atoms described by the Hamiltonian
\begin{eqnarray}
\hat{H}_{a,2}=\Delta( \hat{\sigma}_{1}^{+}\hat{\sigma}_{1}^{-}+\hat{%
\sigma}_{2}^{+}\hat{\sigma}_{2}^{-}),\label{giant atoms}
\end{eqnarray}
where $\Delta=\omega _{e}-\omega _{a}$ is the frequency detuning between the giant atoms and the cavity field. The atomic operators in Eq.~(\ref{giant atoms}) are defined by $\hat{\sigma}_{s=1,2}^{+}=\left\vert e\right\rangle_{ss}\left\langle g\right\vert$ and $\hat{\sigma}_{s}^{-}=\left\vert g\right\rangle_{ss}\left\langle e\right\vert$. Here, $\left\vert g\right\rangle_{s}$ and $\left\vert e\right\rangle_{s}$ are the ground state and excited state of the giant atom $s$, respectively.

According to the coupling-point distributions in the separate-coupling case, there exist sixteen coupling configurations between the atom and the waveguide~\cite{Luo2024}: AAAA, AAAB, AABA, ABAA, BAAA, AABB, ABAB, ABBA, BAAB, BABA, BBAA, ABBB, BABB, BBAB, BBBA, and BBBB coupling distributions. Under the RWA, the interaction Hamiltonian takes the form
\begin{eqnarray}
\hat{H}_{I,2}&=&g[( \mu _{1}\hat{a}_{n_{1}}^{\dag }+\nu _{1}\hat{b}_{n_{1}}^{\dag
}+\mu _{2}\hat{a}_{m_{1}}^{\dag }+\nu _{2}\hat{b}_{m_{1}}^{\dag })
\hat{\sigma}_{1}^{-} \nonumber \\
&&+( \mu _{3}\hat{a}_{n_{2}}^{\dag }+\nu _{3}\hat{b}_{n_{2}}^{\dag
}+\mu _{4}\hat{a}_{m_{2}}^{\dag }+\nu _{4}\hat{b}_{m_{2}}^{\dag })
\hat{\sigma}_{2}^{-}]+\text{H.c.}. \nonumber \\
\label{the interaction Hamiltonian of two giant atoms}
\end{eqnarray}
Here, we assume that the giant atom 1(2) is coupled to the SSH waveguide at both the $n_{1(2)}$th and $m_{1(2)}$th unit cells. For simplicity, we consider that all the coupling points have the same coupling strength $g$. To discuss the photon scattering for different coupling configurations, in Eq.~(\ref{the interaction Hamiltonian of two giant atoms}) we introduce the dimensionless parameters $\mu_{i}$ (for $i=1$--$4$) and $\nu_{i}$ to characterize the coupling between the two giant atoms and the SSH waveguide, where the coefficients $\mu_{i}$ and $\nu_{i}$ take either $0$ or $1$. For example, for the ABAB-coupling configuration, we choose $\mu _{1(3)}=\nu _{2(4)}=1$ and $\mu _{2(4)}=\nu _{1(3)}=0$.

\subsection{Scattering probability amplitudes}\label{Scattering probability amplitudes}
Since the excitation number operator $\hat{N}_{2}=\sum_{s=1}^{2}\hat{\sigma}^{+}_{s}\hat{\sigma}^{-}_{s}+\sum_{j=1}^{N}(\hat{a}_{j}^{\dagger}\hat{a}_{j}+\hat{b}_{j}^{\dagger }\hat{b}_{j})$ of the system is a conserved quantity, a general state in the single-excitation subspace of the system can be written as
\begin{eqnarray}
\left\vert \Psi _{k}\right\rangle _{2}&=&c_{e1}\left\vert e\right\rangle
_{1}\left\vert g\right\rangle _{2}\left\vert \emptyset\right\rangle
+c_{e2}\left\vert g\right\rangle _{1}\left\vert e\right\rangle
_{2}\left\vert \emptyset\right\rangle \nonumber \\
&&+\sum_{j=1}^{N}[ u_{k}(j) \hat{a}_{j}^{\dagger
}+w_{k}(j) \hat{b}_{j}^{\dagger }] \left\vert
g\right\rangle_{1}\left\vert g\right\rangle_{2} \left\vert \emptyset\right\rangle,
\end{eqnarray}
where $c_{e1}$ $(c_{e2})$ is the probability amplitude corresponding to the single excitation stored in the giant atom 1(2), and $u_{k}(j)$ $[w_{k}(j)]$ has been introduced in Eq.~(\ref{eigenstate1}). Similar to Sec.~\ref{Single giant atom}, we can also obtain equations of motion for the probability amplitudes based on the stationary Schr\"{o}dinger equation $\hat{H}_{\text{rot},2}\left\vert \Psi _{k}\right\rangle_{2}=\omega^{(\pm)}\left\vert \Psi _{k}\right\rangle_{2}$. In the case of $j<n_{1}$, $n_{1}<j<m_{1}$, $m_{1}<j<n_{2}$, $n_{2}<j<m_{2}$, and $j>m_{2}$, the probability amplitudes $u_{k}(j)$ and $w_{k}(j)$ corresponding to the upper and lower bands can be assumed to be
\begin{subequations}
\begin{align}
u_{k}(j) &=
\begin{cases}
e^{ikj} + r_{2} e^{-ikj}, & j < n_{1}, \\
A_{1} e^{ikj} + B_{1} e^{-ikj}, & n_{1} < j < m_{1}, \\
A_{2} e^{ikj} + B_{2} e^{-ikj}, & m_{1} < j < n_{2}, \\
A_{3} e^{ikj} + B_{3} e^{-ikj}, & n_{2} < j < m_{2}, \\
t_{2} e^{ikj}, & j > m_{2},
\end{cases}
\end{align}
\begin{align}
w_{k}(j) &=
\begin{cases}
\pm e^{-i\phi_{k}} e^{ikj} \pm e^{i\phi_{k}} r_{2} e^{-ikj}, & j < n_{1}, \\
\pm e^{-i\phi_{k}} A_{1} e^{ikj} \pm e^{i\phi_{k}} B_{1} e^{-ikj}, & n_{1} < j < m_{1}, \\
\pm e^{-i\phi_{k}} A_{2} e^{ikj} \pm e^{i\phi_{k}} B_{2} e^{-ikj}, & m_{1} < j < n_{2}, \\
\pm e^{-i\phi_{k}} A_{3} e^{ikj} \pm e^{i\phi_{k}} B_{3} e^{-ikj}, & n_{2} < j < m_{2}, \\
\pm e^{-i\phi_{k}} t_{2} e^{ikj}, & j > m_{2}.
\end{cases}
\end{align}
\end{subequations}
where $r_{2}$ and $t_{2}$ are the reflection and transmission amplitudes for single photons in the SSH waveguide coupled to two giant atoms, respectively. The variables $A_{1-3}$ and $B_{1-3}$ are the probability amplitudes related to the right- and left-propagating waves within these regions in the waveguide determined by the two giant atoms.

In terms of the continuity conditions at all coupling points, we obtain the relations
\begin{subequations}
\begin{eqnarray}
A_{1}&=&1+\frac{\mu _{1}\pm \nu _{1}e^{i\phi _{k}}}{\mu _{1}\pm \nu_{1}e^{-i\phi _{k}}}(r_{2}-B_{1})e^{-2ikn_{1}}, \\
A_{2}&=&A_{1}+\frac{\mu _{2}\pm \nu _{2}e^{i\phi _{k}}}{\mu _{2}\pm \nu_{2}e^{-i\phi _{k}}}(B_{1}-B_{2})e^{-2ikm_{1}}, \\
A_{3}&=&A_{2}+\frac{\mu _{3}\pm \nu _{3}e^{i\phi _{k}}}{\mu _{3}\pm \nu_{3}e^{-i\phi _{k}}}(B_{2}-B_{3})e^{-2ikn_{2}}, \\
t_{2}&=&A_{3}+\frac{\mu _{4}\pm \nu _{4}e^{i\phi _{k}}}{\mu _{4}\pm\nu _{4}e^{-i\phi _{k}}}B_{3}e^{-2ikm_{2}}.
\end{eqnarray}
\end{subequations}

\begin{widetext}
Then, the unified expressions of the transmission and reflection amplitudes for all coupling configurations can be obtained as
\begin{subequations}
\begin{eqnarray}
t_{2}^{(\pm )}\!\!&=\!\!&\frac{\left(i\Delta _{k1}^{(\pm )}-i\Delta _{L1}^{(\pm)}\right)(1\leftrightarrow 2)}{\left(i\Delta _{k1}^{(\pm )}-i\Delta _{L1}^{(\pm )}-\Gamma _{1}^{(\pm )}/2 \right)(1\leftrightarrow 2)-\left(\Gamma _{12}^{(\pm )}/2+ig_{12}^{(\pm )} \right)^{2}},\label{t2} \\
r_{2}^{(\pm )}\!\!&=\!\!&\frac{\Gamma _{1}^{(\pm )}\left(i\Delta _{k2}^{(\pm )}-i\Delta _{L2}^{(\pm
)}-\Gamma _{2}^{(\pm )}/2 \right)e^{2ikn_{1}+i\theta _{1}^{(\pm)}}\!\!+(1\leftrightarrow 2)+\Gamma _{1}^{(\pm )}\Gamma _{2}^{(\pm)}e^{2ikn_{2}+i\theta _{2}^{(\pm )}}}{2 \left[\left(i\Delta _{k1}^{(\pm )}-i\Delta
_{L1}^{(\pm )}-\Gamma _{1}^{(\pm )}/2 \right)(1\leftrightarrow 2)-\left(\Gamma _{12}^{(\pm )}/2+ig_{12}^{(\pm )} \right)^{2}\right]}, \label{r2}
\end{eqnarray}
\end{subequations}
where $\Delta _{k1}^{(\pm)}=\Delta _{k2}^{(\pm)}=\Delta _{k}^{(\pm)}=\pm\omega_{k}-\Delta$ are the detunings between the upper and lower band propagating photons in the SSH waveguide and the giant atoms. Note that the notation ($1\leftrightarrow 2$) in Eqs.~(\ref{t2}) and (\ref{r2}) represents the term having the same form as the former term under the replacement of the subscripts $1\leftrightarrow 2$. We can check that the transmission and reflection amplitudes satisfy the relation $|t_{2}^{(\pm )}|^{2}+|r_{2}^{(\pm )}|^{2}=1$. It can be seen that the scattering amplitudes are determined by four characteristic quantities: the Lamb shifts $\Delta _{L1(2)}^{(\pm)}$, the individual decay rates $\Gamma _{1(2)}^{(\pm)}$ of the giant atom 1(2), the exchange interaction strengths $g_{12}^{(\pm)}$, and the collective decay rates $\Gamma _{12}^{(\pm)}$ of the two giant atoms, which are defined as
\begin{subequations}
\begin{eqnarray}
\Delta _{L1}^{(\pm )} &=&\Gamma_{e}^{(\pm )}[(\mu _{1}\mu _{2}+\nu
_{1}\nu _{2})\sin (kd_{1})\mp (\mu _{1}\nu _{1}+\mu _{2}\nu _{2})\sin \phi
_{k}  \nonumber \\
&&\pm \nu _{1}\mu _{2}\sin (kd_{1}+\phi _{k})\pm \mu _{1}\nu _{2}\sin
(kd_{1}-\phi _{k})],  \label{Lamb shift 1} \\
\Delta _{L2}^{(\pm )} &=&\Gamma_{e}^{(\pm )}[(\mu _{3}\mu _{4}+\nu
_{3}\nu _{4})\sin (kd_{2})\mp (\mu _{3}\nu _{3}+\mu _{4}\nu _{4})\sin \phi
_{k}  \nonumber \\
&&\pm \nu _{3}\mu _{4}\sin (kd_{2}+\phi _{k})\pm \mu _{3}\nu _{4}\sin
(kd_{2}-\phi _{k})],  \label{Lamb shift 2} \\
\Gamma _{1}^{(\pm )} &=&2\Gamma_{e}^{(\pm )}[\pm (\mu _{1}\nu _{1}+\mu _{2}\nu _{2})\cos \phi _{k}+(\mu _{1}\mu
_{2}+\nu _{1}\nu _{2})\cos (kd_{1}) \nonumber \\
&&\pm \mu _{1}\nu_{2}\cos (kd_{1}-\phi _{k})\pm \nu _{1}\mu _{2}\cos (kd_{1}+\phi _{k})+1],\label{individual decay 1} \\
\Gamma _{2}^{(\pm )} &=&2\Gamma_{e}^{(\pm )}[\pm (\mu _{3}\nu _{3}+\mu _{4}\nu _{4})\cos \phi _{k}+(\mu _{3}\mu
_{4}+\nu _{3}\nu _{4})\cos (kd_{2}) \nonumber \\
&&\pm \mu _{3}\nu_{4}\cos (kd_{2}-\phi _{k})\pm \nu _{3}\mu _{4}\cos (kd_{2}+\phi _{k})+1],\label{individual decay 2}
\end{eqnarray}
%%%%%%%%%%%%%%%%%%%%%%%%%%%
\begin{table*}[ht]
    \renewcommand{\arraystretch}{1.1}
    \caption{The relationships between the reflection coefficients for different coupling configurations. Here, the superscript $\text{C}_{1}\text{C}_{2}\text{C}_{3}\text{C}_{4}$ denotes the coupling configuration with $\text{C}_{i}\in\{\text{A},\text{B}\}$ (for $i=1$--$4$) and the subscript $\pm$ $(\mp)$ represents the sign of the dimerization parameter $\delta$.}
    \label{table_two}
    \begin{ruledtabular}
        \begin{tabular*}{\linewidth}{ccccc}
        \makecell[l]{Coupling\\ configurations} & $d_{1}$ & $d_{2}$ & $d_{21}$ & \makecell[c]{Reflection\\ coefficients} \\
        \hline
        AAAA & $x_{1}^{\text{AAAA}}$ & $x_{2}^{\text{AAAA}}$ & $x_{21}^{\text{AAAA}}$ & $R_{2,\pm}^{\text{AAAA}}=R_{2,\pm}^{\text{BBBB}}$ \\
        BBBB & $x_{1}^{\text{AAAA}}$ & $x_{2}^{\text{AAAA}}$ & $x_{21}^{\text{AAAA}}$ & \\
        AAAB & $x_{1}^{\text{AAAB}}$ & $x_{2}^{\text{AAAB}}$ & $x_{21}^{\text{AAAB}}$ & $R_{2,\pm}^{\text{AAAB}}=R_{2,\pm}^{\text{ABBB}}=R_{2,\mp}^{\text{BAAA}}=R_{2,\mp}^{\text{BBBA}}$ \\
        ABBB & $x_{2}^{\text{AAAB}}$ & $x_{1}^{\text{AAAB}}$ & $x_{21}^{\text{AAAB}}$ & \\
        BAAA & $x_{2}^{\text{AAAB}}+1$ & $x_{1}^{\text{AAAB}}$ & $x_{21}^{\text{AAAB}}$ & \\
        BBBA & $x_{1}^{\text{AAAB}}$ & $x_{2}^{\text{AAAB}}+1$ & $x_{21}^{\text{AAAB}}$ & \\
        AABA & $x_{1}^{\text{AABA}}$ & $x_{2}^{\text{AABA}}$ & $x_{21}^{\text{AABA}}$ & $R_{2,\pm}^{\text{AABA}}=R_{2,\pm}^{\text{BABB}}=R_{2,\mp}^{\text{ABAA}}=R_{2,\mp}^{\text{BBAB}}$ \\
        BABB & $x_{2}^{\text{AABA}}$ & $x_{1}^{\text{AABA}}$ & $x_{21}^{\text{AABA}}$ & \\
        ABAA & $x_{2}^{\text{AABA}}-1$ & $x_{1}^{\text{AABA}}$ & $x_{21}^{\text{AABA}}+1$ & \\
        BBAB & $x_{1}^{\text{AABA}}$ & $x_{2}^{\text{AABA}}-1$ & $x_{21}^{\text{AABA}}+1$ & \\
        BAAB & $x_{1}^{\text{BAAB}}$ & $x_{2}^{\text{BAAB}}$ & $x_{21}^{\text{BAAB}}$ & $R_{2,\pm}^{\text{BAAB}}=R_{2,\pm}^{\text{ABBA}}$ \\
        ABBA & $x_{2}^{\text{BAAB}}$ & $x_{1}^{\text{BAAB}}$ & $x_{21}^{\text{BAAB}}$ & \\
        AABB & $x_{1}^{\text{AABB}}$ & $x_{2}^{\text{AABB}}$ & $x_{21}^{\text{AABB}}$ & $R_{2,\pm}^{\text{AABB}}=R_{2,\mp}^{\text{BBAA}}$ \\
        BBAA & $x_{2}^{\text{AABB}}$ & $x_{1}^{\text{AABB}}$ & $x_{21}^{\text{AABB}}+1$ & \\
        ABAB & $x_{1}^{\text{ABAB}}$ & $x_{2}^{\text{ABAB}}$ & $x_{21}^{\text{ABAB}}$ & $R_{2,\pm}^{\text{ABAB}}=R_{2,\mp}^{\text{BABA}}$ \\
        BABA & $x_{1}^{\text{ABAB}}+1$ & $x_{2}^{\text{ABAB}}+1$ & $x_{21}^{\text{ABAB}}-1$ & \\
        \end{tabular*}
    \end{ruledtabular}
    \label{table:The relationship of transmission amplitudes}
\end{table*}
%%%%%%%%%%%%%%%%%%%%%%%%%%%%%%%%
\begin{eqnarray}
\Gamma _{12}^{(\pm )} &=&\Gamma_{e}^{(\pm )}\{\pm \mu _{2}\mu
_{3}\cos kd_{21}\pm \mu _{1}\mu _{3}\cos k(d_{21}+d_{1})  \nonumber \\
&&\pm \mu _{2}\mu _{4}\cos k(d_{21}+d_{2})\pm \mu _{1}\mu _{4}\cos
k(d_{21}+d_{1}+d_{2})  \nonumber \\
&&+\mu _{2}\nu _{3}\cos (kd_{21}+\phi _{\mu })+\mu _{1}\nu _{3}\cos
[k(d_{21}+d_{1})+\phi _{\mu }]  \nonumber \\
&&+\mu _{1}\nu _{4}\cos [k(d_{21}+d_{1}+d_{2})+\phi _{\mu }]  \nonumber \\
&&+\mu _{2}\nu _{4}\cos [k(d_{21}+d_{2})+\phi _{\mu }]\}+\Gamma_{e}^{(\pm )}(\mu \leftrightarrow \nu) , \label{collective decay} \\
g_{12}^{(\pm )} &=&\Gamma_{e}^{(\pm )}\{\pm \mu _{2}\mu _{3}\sin
kd_{21}\pm \mu _{1}\mu _{3}\sin k(d_{21}+d_{1})  \nonumber \\
&&\pm \mu _{2}\mu _{4}\sin k(d_{21}+d_{2})\pm \mu _{1}\mu _{4}\sin
k(d_{21}+d_{1}+d_{2})  \nonumber \\
&&+\mu _{2}\nu _{3}\sin (kd_{21}+\phi _{\mu })+\mu _{1}\nu _{3}\sin
[k(d_{21}+d_{1})+\phi _{\mu }]  \nonumber \\
&&+\mu _{1}\nu _{4}\sin [k(d_{21}+d_{1}+d_{2})+\phi _{\mu }]  \nonumber \\
&&+\mu _{2}\nu _{4}\sin [k(d_{21}+d_{2})+\phi _{\mu }]\}/2+\Gamma_{e}^{(\pm )}(\mu \leftrightarrow \nu)/2, \label{exchange interaction} \\
\theta _{1}^{(\pm )}&=&\text{Arg}\left[\Gamma_{e}^{(\pm )}[(\mu _{1}\pm \nu
_{1}e^{-i\phi _{k}})+(1\leftrightarrow 2)e^{ikd_{1}}]^{2}\right],\label{phase 1} \\
\theta _{2}^{(\pm )}&=&\text{Arg}\left[ \Gamma_{e}^{(\pm )}[(\mu _{3}\pm \nu
_{3}e^{-i\phi _{k}})+(3\leftrightarrow 4)e^{ikd_{2}}]^{2}\right]. \label{phase 2}
\end{eqnarray}
\end{subequations}
\end{widetext}

In Eqs.~(\ref{Lamb shift 1})--(\ref{phase 2}), $d_{1(2)}=m_{1(2)}-n_{1(2)}$ is the coupling-point distance of the giant atom 1(2) and $d_{21}=n_{2}-m_{1}$ is the distance between the second coupling point of the giant atom 1 and the first coupling point of the giant atom 2. To get more compact expressions, we have introduced $\phi_{\mu}=-\phi_{\nu}=-\phi_{k}$ in Eqs.~(\ref{collective decay}) and (\ref{exchange interaction}). In Eqs.~(\ref{t2})--(\ref{exchange interaction}), we consider that the incident photon frequency is $\omega_{k}$ with $k\in(-\pi,0)$ and takes different values of $\mu_{i}$ and $\nu_{i}$ for $i=1$--$4$. We can obtain the concrete forms of the reflection coefficients for 16 coupling configurations. We find that, for the AAAA- and BBBB-coupling configurations, the reflection coefficients are only related to the coupling-point distances. For other coupling configurations, the reflection coefficients also depend on the topology-dependent phase $\phi_{k}$. The relations of these reflection coefficients correspond to different configurations and are listed in Table~\ref{table:The relationship of transmission amplitudes}. By changing the value of $d$ and the sign of $\delta$, these 16 coupling configurations can be reduced into six coupling cases: AAAA, AAAB, AABA, ABBA, AABB, and ABAB couplings. To explain the physical mechanism behind this phenomenon, we introduce the symmetric-antisymmetric (S-A) lowering operators $\hat{\sigma}_{\text{S,A}}^{-}=(\hat{\sigma}_{1}^{-}\pm\hat{\sigma}_{2}^{-})/\sqrt 2$~\cite{Bello2019}. In this case, the two two-level giant atoms in the single-excitation subspace can be understood as a V-type three-level atom coupled to the SSH waveguide through the symmetric- and antisymmetric-state channels. For different coupling configurations, the symmetric and antisymmetric states are coupled to the SSH waveguide with different coupling strengths. In the S-A state representation, we introduce the characteristic parameters as
\begin{subequations}
\begin{eqnarray}
\Delta _{\text{S(A)}}&=&\Delta _{k}-\Delta _{\text{S(A)}L}, \label{effective detuning S A} \\
\Delta _{\text{S(A)}L}&=&(\Delta _{L1}+\Delta _{L2}\pm 2g_{12})/2, \label{Lamb shift S A}
\end{eqnarray}
\begin{eqnarray}
\Gamma _{\text{S(A)}}&=&( \Gamma _{1}+\Gamma _{2}\pm 2\Gamma_{12})/2,\label{individual decay S A} \\
g_{\text{SA}}&=&(\Delta _{L1}-\Delta_{L2})/2,\label{exchange interaction SA} \\
\Gamma _{\text{SA}}&=&( \Gamma _{1}-\Gamma _{2})/2.\label{collective decay SA}
\end{eqnarray}
\end{subequations}
Here, $\Delta _{\text{S(A)}}$ is the effective detuning between the propagating photons in the waveguide and the symmetric (antisymmetric) state, $\Gamma _{\text{S(A)}}$ is the individual decay rate of the symmetric (antisymmetric) state, and $g_{\text{SA}}$ and $\Gamma_{\text{SA}}$ are the exchange interaction strength and collective decay rate, respectively. According to Eqs.~(\ref{effective detuning S A})--(\ref{collective decay SA}), the transmission and reflection amplitudes in Eqs.~(\ref{t2}) and (\ref{r2}) can be reexpressed by $\Delta _{\text{S(A)}}$, $\Delta _{\text{S(A)}L}$, $\Gamma _{\text{S(A)}}$, $g_{\text{SA}}$, and $\Gamma _{\text{SA}}$.

\subsection{Scattering spectra}\label{Scattering spectra2}
Based on the relationship indicated by the reflection coefficients in Table~\ref{table:The relationship of transmission amplitudes}, we only need to analyze the scattering spectra of these six coupling configurations. In the following, we investigate the influence of the coupling configurations, frequency detuning, and sign of the dimerization parameter on the single-photon scattering in the weak-coupling regime. Similar to the method in Sec.~\ref{Single giant atom}, we can replace the wave vector $k$ in Eqs.~(\ref{effective detuning S A})--(\ref{collective decay SA}) with $k_{\Delta}$. Note that the single-photon scattering is also periodically modulated by the coupling-point distance (with a period of 4). Therefore, we take fixed values of $d$ in the following discussions.

%%%%%%%%%%%%%%%%%%%%%%%%%%%%%
\begin{figure}[tbp]
\center\includegraphics[width=0.48\textwidth]{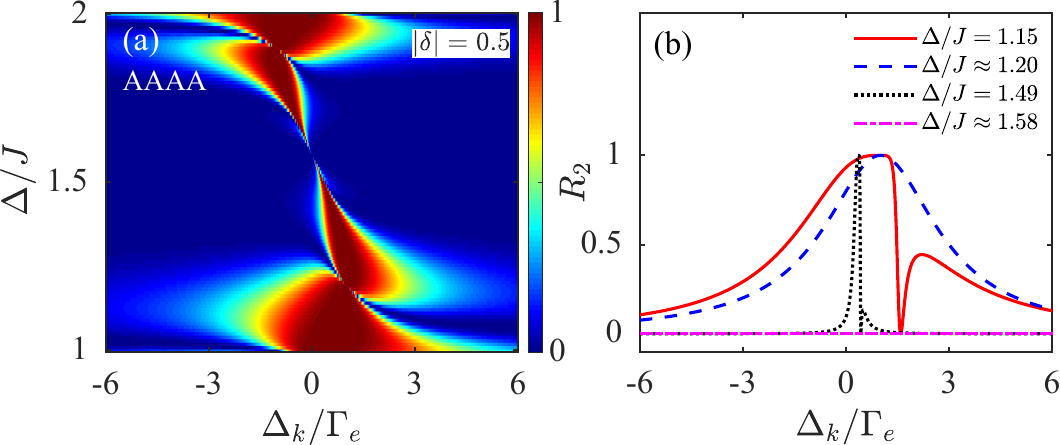}
\caption{Reflection coefficient $R_{2}$ for the AAAA-coupling configuration as a function of $\Delta _{k}$ and $\Delta$. The dimerization parameter is (a) $|\delta|=0.5$. The profiles of panel (a) at several values of $\Delta$ are shown in (b). Other parameters used are $g=0.01J$ and $d=2$.}
\label{RAAAAvsDeltakchgDelta}
\end{figure}
%%%%%%%%%%%%%%%%%%%%%%%%%%%%%
\subsubsection{The AAAA-coupling configuration}
For the AAAA-coupling configuration, we have $\mu_{i}=1$ and $\nu_{i}=0$ (for $i=1$--$4$) in Eqs.~(\ref{t2})--(\ref{exchange interaction}). Figure~\ref{RAAAAvsDeltakchgDelta}(a) shows the reflection coefficient as a function of $\Delta_{k}$ and $\Delta$. Similar to the AA-coupling configuration in the single-giant-atom case, the sign of $\delta$ has no influence on the single-photon scattering behavior for the AAAA-coupling configuration and we only need to analyze the influence of $|\delta|$ on the scattering spectra. However, the reflection spectra show more abundant line shapes due to the collective effect of the two giant atoms.

In Fig.~\ref{RAAAAvsDeltakchgDelta}(b), we plot the profile of Fig.~\ref{RAAAAvsDeltakchgDelta}(a) at several frequency detunings in the region of $\Delta\in(J,2J)$; we find that when $\Delta\approx1.20J$ (the blue dashed curve), the characteristic quantities in Eqs.~(\ref{effective detuning S A})--(\ref{collective decay SA}) become $\Delta _{\text{S}L,k_{\Delta}}=\Delta _{\text{A}L,k_{\Delta}}=\Gamma_{e}$, $\Gamma _{\text{S},k_{\Delta}}=0$, $\Gamma _{\text{A},k_{\Delta}}=4\Gamma_{e}$, and $g_{\text{SA},k_{\Delta}}=\Gamma_{\text{SA},k_{\Delta}}=0$. Therefore, the symmetric (antisymmetric) state is decoupled from (coupled to) the waveguide. This feature results in the appearance of the Lorentzian line shape centered at $\Delta _{k}=\Gamma _{e}$ with the linewidth of $4\Gamma _{e}$. When $\Delta\approx1.58J$ (the purple dot-dashed curve), the six characteristic quantities are $\Delta _{\text{S}L,k_{\Delta}}=\Delta _{\text{A}L,k_{\Delta}}=\Gamma _{\text{S},k_{\Delta}}=\Gamma_{\text{A},k_{\Delta}}=g_{\text{SA},k_{\Delta}}=\Gamma_{\text{SA},k_{\Delta}}=0$. In this case, both the symmetric and antisymmetric states are decoupled from the SSH waveguide. Therefore, the incident photon in the waveguide can be completely transmitted. When $\Delta=1.15J$ (the red solid curve) and $1.49J$ (the black dotted curve), the spectra become asymmetric, with a minimal reflection located at $\Delta _{k}=\Gamma_{e} \{\sin (k_{\Delta}d)-[1+\cos (k_{\Delta}d)]\tan (2k_{\Delta}d)\}$.

In the following analyses, we can show that in some regions the reflection spectrum around the minimal reflection can be approximated as a standard Fano line shape~\cite{Fano1961,Miroshnichenko2010}. To this end, we decompose the reflection amplitude in Eq.~(\ref{r2}) as $r_{2,\pm}^{\text{AAAA}}=r_{2,+}^{\text{AAAA}}+r_{2,-}^{\text{AAAA}}$, where
\begin{eqnarray}
r_{2,\pm}^{\text{AAAA}}=\frac{\pm\Gamma _{\pm }e^{ik_{\Delta}d}e^{2ik_{\Delta}m_{1}}/2}{i(\Delta
_{k}-\Delta _{\pm })-\Gamma _{\pm }/2}\label{rpm}
\end{eqnarray}
correspond to two Lorentz spectra. In Eq.~(\ref{rpm}), we introduce the resonance points $\Delta _{\pm }$ and the widths $\Gamma _{\pm }$, which are given by
\begin{subequations}
\begin{eqnarray}
\Delta _{\pm }&=&\Gamma_{e} \sin (k_{\Delta}d)[1\pm 2\cos (k_{\Delta}d)\pm 2\cos ^{2}(k_{\Delta}d)],\label{Lamb shift pm} \\
\Gamma _{\pm }&=&2\Gamma_{e} [1+\cos (k_{\Delta}d)][1\pm \cos (2k_{\Delta}d)].\label{individual decay pm}
\end{eqnarray}
\end{subequations}
To obtain the reflection spectra of the Fano line shapes, the widths $\Gamma _{\pm }$ should satisfy the condition $\Gamma _{+}\gg \Gamma _{-}$ or $\Gamma _{-}\gg \Gamma _{+}$~\cite{Feng2021}, which can be realized by adjusting the system parameters. For $\Gamma _{+}\gg \Gamma _{-}$ ($\Gamma _{-}\gg \Gamma _{+}$), the $r_{2,+}$ ($r_{2,-}$) becomes a continuous mode, while the $r_{2,-}$ ($r_{2,+}$) becomes a discrete mode. When the two modes interact with each other, a quantum interference phenomenon will take place. This leads to the appearance of the Fano reflection spectra. In particular, when $\Gamma _{\pm}\gg \Gamma _{\mp}$, the reflection coefficient around $\Delta _{\mp}$ can be approximated as~\cite{Miroshnichenko2010}
\begin{eqnarray}
R_{2}\approx \eta\frac{\left( q+\epsilon \right) ^{2}}{1+\epsilon ^{2}},\label{Fano line shape}
\end{eqnarray}
where we introduce the asymmetry parameter $q=(\Delta _{\pm}-\Delta _{\mp})/(\Gamma _{\pm}/2)$, the reduced detuning  $\epsilon =(\Delta _{k}-\Delta _{\mp})/(\Gamma _{\mp}/2)$, and the modified coefficient $\eta=(\Gamma _{\pm}^{2}/4)/[(\Delta _{\pm}-\Delta _{\mp})^{2}+(\Gamma _{\pm}^{2}/4)]$. Concretely, when the frequency detuning $\Delta$ is taken as $\Delta/J\in(1,1.02)\cup(1.48,1.58)\cup(1.58,1.68)\cup(1.99,2)$, we have $\Gamma _{+}\gg \Gamma _{-}$. In this case, the reflection spectra can exhibit the Fano line shapes around the minimal reflection, as shown by the black dotted curve in Fig.~\ref{RAAAAvsDeltakchgDelta}(b). On the other hand, when $\Delta/J\in(1.11,1.20)\cup(1.20,1.30)\cup(1.82,1.89)\cup(1.89,1.94)$, we have $\Gamma _{-}\gg \Gamma _{+}$. Then the reflection spectra can be approximated as the Fano line shapes, as shown by the red solid curve in Fig.~\ref{RAAAAvsDeltakchgDelta}(b).

%%%%%%%%%%%%%%%%%%%%%%%%%%%%%
\begin{figure}[tbp]
\center\includegraphics[width=0.48\textwidth]{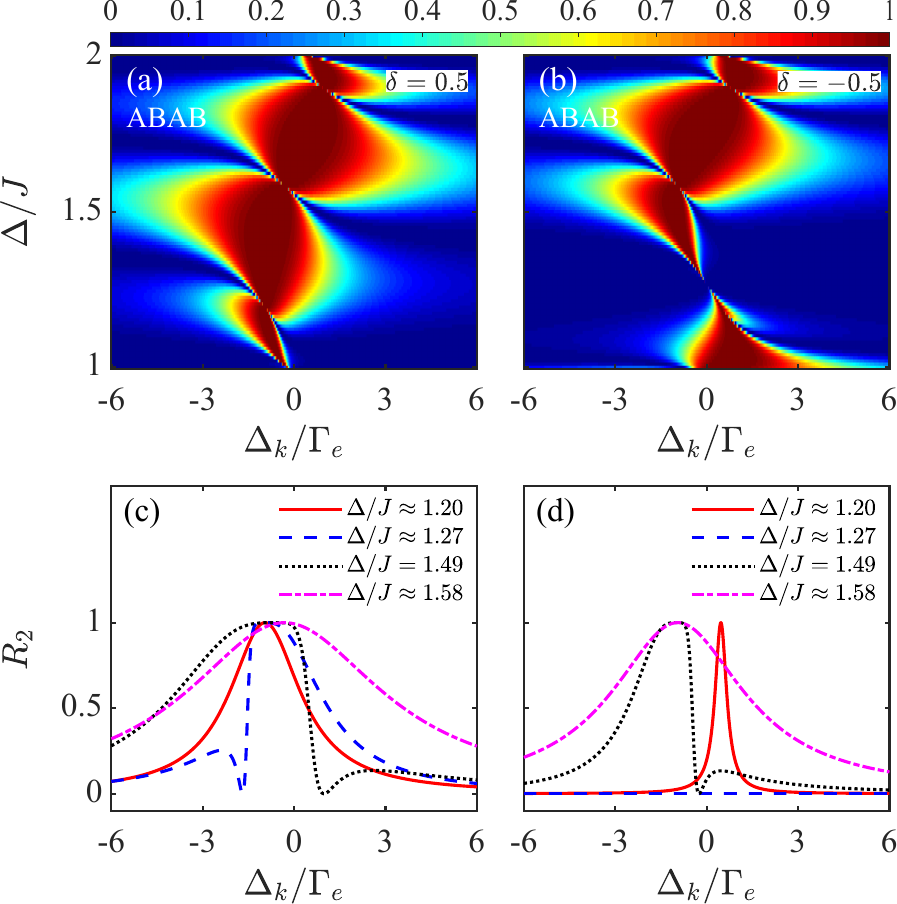}
\caption{Reflection coefficient $R_{2}$ for the ABAB-coupling configuration as a function of $\Delta _{k}$ and $\Delta$. The dimerization parameter is (a) $\delta=0.5$ and (b) $\delta=-0.5$. The profiles of panels (a) and (b) at several values of $\Delta$ are shown by the curves in (c) and (d). Other parameters used are $g=0.01J$ and $d=2$.}
\label{RABABvsDeltakchgDelta}
\end{figure}
%%%%%%%%%%%%%%%%%%%%%%%%%%%%%
\subsubsection{The ABAB-coupling configuration}
For the ABAB-coupling configuration, the dimensionless parameters become $\mu_{1(3)}=\nu_{2(4)}=1$ and $\nu_{1(3)}=\mu_{2(4)}=0$ in Eqs.~(\ref{t2})--(\ref{exchange interaction}). To see the influence of $\Delta$ and $\delta$ on the single-photon scattering process, in Figs.~\ref{RABABvsDeltakchgDelta}(a) and \ref{RABABvsDeltakchgDelta}(b) we show the reflection coefficient as a function of $\Delta _{k}$ and $\Delta$. Since the characteristic quantities contain the topology-dependent phase, it can be seen that the line shapes of the reflection spectra are sensitive to the sign of $\delta$, which is different from the AAAA-coupling configuration. To show the details more clearly, in Figs.~\ref{RABABvsDeltakchgDelta}(c) and \ref{RABABvsDeltakchgDelta}(d), we plot the profiles of Figs.~\ref{RABABvsDeltakchgDelta}(a) and \ref{RABABvsDeltakchgDelta}(b) at different frequency detunings in the region $\Delta \in(J,2J)$, respectively.

We can see from the red solid curve in Fig.~\ref{RABABvsDeltakchgDelta}(c) [Fig.~\ref{RABABvsDeltakchgDelta}(d)] that, when $\delta=0.5$ ($-0.5$) and $\Delta\approx1.20J$, the characteristic quantities are $\Delta _{\text{S}L,k_{\Delta}}=\Delta _{\text{A}L,k_{\Delta}}\approx -0.96\Gamma_{e}$ ($0.47\Gamma_{e}$), $\Gamma _{\text{S},k_{\Delta}}\approx 0$, $\Gamma _{\text{A}k_{\Delta}}\approx 2.82\Gamma_{e}$ ($0.46\Gamma_{e}$), and $g_{\text{SA},k_{\Delta}}=\Gamma_{\text{SA},k_{\Delta}}=0$. Therefore, the symmetric (antisymmetric) state is decoupled from (coupled to) the SSH waveguide. This makes the reflection spectrum exhibit the Lorentzian line shape centered at $\Delta _{k}=-0.96\Gamma _{e}$ ($0.47\Gamma_{e}$) with the linewidth of $2.82\Gamma _{e}$ ($0.46\Gamma_{e}$), as shown by the red solid curve. However, when $\delta=0.5$ ($-0.5$) and $\Delta\approx1.58J$, i.e., the purple dot-dashed curve, the characteristic quantities become $\Delta _{\text{S}L,k_{\Delta}}=\Delta _{\text{A}L,k_{\Delta}}\approx -0.32\Gamma_{e}$ ($-0.95\Gamma_{e}$), $\Gamma _{\text{S},k_{\Delta}}\approx 7.79\Gamma_{e}$ ($5.26\Gamma_{e}$), $\Gamma _{\text{A}k_{\Delta}}\approx 0$, and $g_{\text{SA},k_{\Delta}}=\Gamma_{\text{SA},k_{\Delta}}=0$. In this case, the antisymmetric (symmetric) state is decoupled from (coupled to) the SSH waveguide, and then the reflection spectrum also exhibits the Lorentzian line shape but centered at $\Delta _{k}=-0.32\Gamma _{e}$ ($-0.95\Gamma_{e}$) with the linewidth of $7.79\Gamma _{e}$ ($5.26\Gamma_{e}$), as shown by the purple dot-dashed curve. When $\delta=\pm0.5$ and $\Delta=1.49J$ (the black dotted curves), the spectra around the minimal reflection can be approximated as the asymmetric Fano line shapes. Here, the reflection spectrum takes the minima value ($R_{2}=0$) at $\Delta _{k}=\Gamma_{e} \{\sin (k_{\Delta}d-\phi_{k_{\Delta}})-[1+\cos (k_{\Delta}d-\phi_{k_{\Delta}})]\tan (2k_{\Delta}d) \}$. Specially, when $\delta=\pm0.5$ and $\Delta\approx1.27J$ (the blue dashed curves), the reflection spectra show the asymmetric Fano line shape and complete transmission, respectively. Similar to the AAAA-coupling configuration, we rewrite the reflection amplitude given by Eq.~(\ref{r2}) as the superposition of two Lorentz spectra $r_{2,\pm}^{\text{ABAB}}=r_{2,+}^{\text{ABAB}}+r_{2,-}^{\text{ABAB}}$, where the Lorentz-type amplitudes are given by
\begin{eqnarray}
r_{2}^{\text{ABAB}}=\frac{\pm\Gamma _{\pm }e^{i(k_{\Delta}d-\phi_{k_{\Delta}})}e^{2ik_{\Delta}m_{1}}/2}{i(\Delta
_{k}-\Delta _{\pm })-\Gamma _{\pm }/2},\label{ABAB rpm}
\end{eqnarray}
with
\begin{subequations}
\begin{eqnarray}
\Delta _{\pm }&=&\Gamma_{e}\{\sin(k_{\Delta}d-\phi_{k_{\Delta}})\pm[1+\cos(k_{\Delta}d-\phi_{k_{\Delta}})]\sin (2k_{\Delta}d) \}, \nonumber \\
&& \label{ABAB Lamb shift pm} \\
\Gamma _{\pm }&=&2\Gamma_{e}[1+\cos(k_{\Delta}d-\phi_{k_{\Delta}})][1\pm \cos (2k_{\Delta}d)].\label{ABAB individual decay pm}
\end{eqnarray}
\end{subequations}

%%%%%%%%%%%%%%%%%%%%%%%%%%%%%
\begin{figure}[tbp]
\center\includegraphics[width=0.48\textwidth]{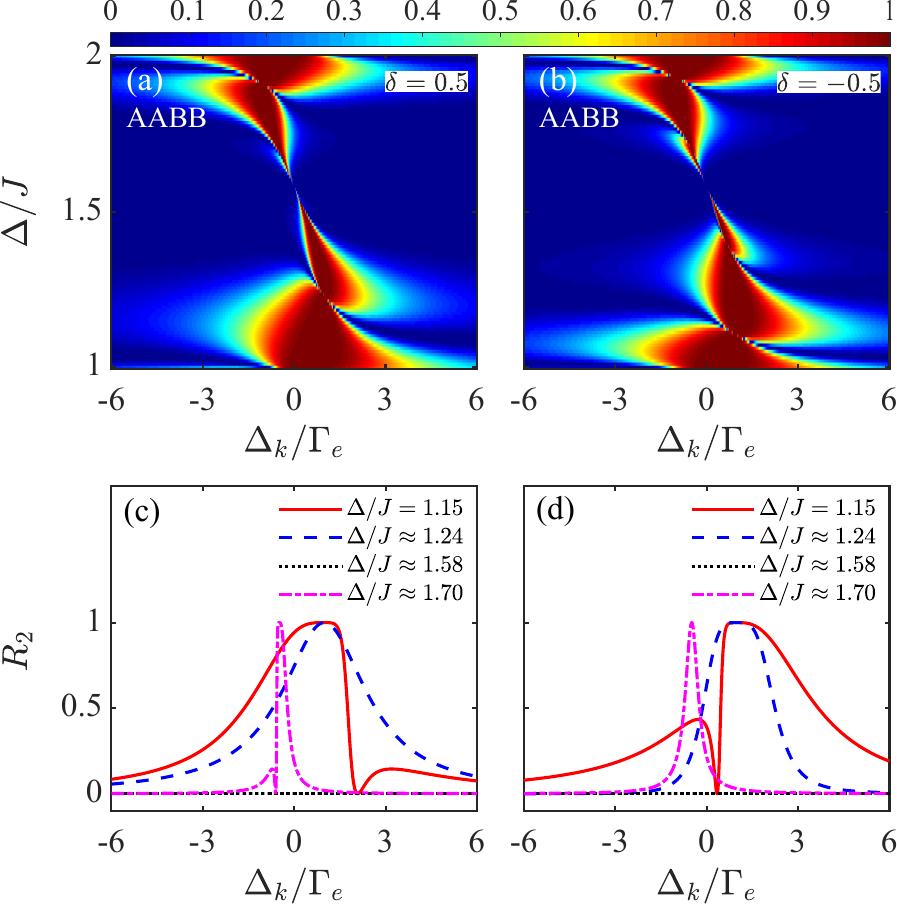}
\caption{Reflection coefficient $R_{2}$ for the AABB-coupling configuration as a function of $\Delta _{k}$ and $\Delta$. The dimerization parameter is (a) $\delta=0.5$ and (b) $\delta=-0.5$. The profiles of panels (a) and (b) at several values of $\Delta$ are shown by the curves in (c) and (d). Other parameters used are $g=0.01J$ and $d=2$.}
\label{RAABBvsDeltakchgDelta}
\end{figure}
%%%%%%%%%%%%%%%%%%%%%%%%%%%%%
To obtain the Fano line shapes for the reflection spectra, we need to ensure the condition $\Gamma _{\pm}\gg \Gamma _{\mp}$. Under this condition, the reflection spectrum near $\Delta _{\mp}$ can be approximated as the Fano line shape characterized by Eq.~(\ref{Fano line shape}). Specifically, when the dimerization parameters $\delta=\pm0.5$ and $\Delta$ is taken as $\Delta/J\in(1,1.02)\cup(1.48,1.58)\cup(1.58,1.68)\cup(1.99,2)$, the width $\Gamma _{\pm}$ satisfies $\Gamma _{+}\gg \Gamma _{-}$. Thus, the reflection spectra exhibit the Fano line shapes around the reflection minima, as shown by the black dotted curve in Figs.~\ref{RABABvsDeltakchgDelta}(c) and \ref{RABABvsDeltakchgDelta}(d). When the dimerization parameters $\delta=\pm0.5$ and $\Delta/J\in(1.11,1.20)\cup(1.20,1.27)\cup(1.27,1.30)\cup(1.82,1.89)\cup(1.89,1.94)$, we have $\Gamma _{-}\gg \Gamma _{+}$, and then the reflection spectrum exhibits the Fano line shape, as shown by the blue dashed curve in Fig.~\ref{RABABvsDeltakchgDelta}(c) (i.e., $\delta=0.5$ and $\Delta\approx1.27J$). When $\delta=-0.5$ and $\Delta\approx1.27J$, the characteristic quantities in Eqs.~(\ref{effective detuning S A})--(\ref{collective decay SA}) become $\Delta _{\text{S}L,k_{\Delta}}=\Delta _{\text{A}L,k_{\Delta}}=\Gamma _{\text{S},k_{\Delta}}=\Gamma_{\text{A},k_{\Delta}}=g_{\text{SA},k_{\Delta}}=\Gamma_{\text{SA},k_{\Delta}}=0$. Therefore, both the symmetric and antisymmetric states are decoupled from the SSH waveguide. In this case, the input photon can be completely transmitted by the waveguide, as shown by the blue dashed curve in Fig.~\ref{RABABvsDeltakchgDelta}(d).

\subsubsection{The AABB-coupling configuration}
For the AABB-coupling configuration, the dimensionless parameters in Eqs.~(\ref{t2})--(\ref{exchange interaction}) are $\mu_{1(2)}=\nu_{3(4)}=1$ and $\nu_{1(2)}=\mu_{3(4)}=0$. Figures~\ref{RAABBvsDeltakchgDelta}(a) and \ref{RAABBvsDeltakchgDelta}(b) show the reflection coefficient as a function of $\Delta _{k}$ and $\Delta$. We find that the sign of $\delta$ can affect the reflection spectra of the AABB-coupling configuration, but not affect the location of the reflection peak ($R_{2}=1$). This is because the Lamb shifts $\Delta_{L1,k_{\Delta}}=\Delta_{L2,k_{\Delta}}=\Gamma _{e}\sin(k_{\Delta}d)$ are independent of the sign of $\delta$. The feature is different from the ABAB-coupling configuration.

In Figs.~\ref{RAABBvsDeltakchgDelta}(c) and \ref{RAABBvsDeltakchgDelta}(d), we plot the profiles of Figs.~\ref{RAABBvsDeltakchgDelta}(a) and \ref{RAABBvsDeltakchgDelta}(b) at different frequency detunings in the region $\Delta \in(J,2J)$, respectively. It is found that, when $\delta=\pm0.5$ and $\Delta=1.15J$ (the red solid curves), the spectra around the minimal reflection can be approximated as the asymmetric Fano line shapes. Here, the complete transmission ($R_{2}=0$) appears at $\Delta _{k}=\Gamma_{e} \{\sin (k_{\Delta}d)-[1+\cos (k_{\Delta}d)]\tan (2k_{\Delta}d-\phi_{k_{\Delta}})\}$. Then we can rewrite the reflection amplitude Eq.~(\ref{r2}) as the superposition of two Lorentz spectra $r_{2,\pm}^{\text{AABB}}=r_{2,+}^{\text{AABB}}+r_{2,-}^{\text{AABB}}$, where
\begin{eqnarray}
r_{2,\pm}^{\text{AABB}}=\frac{\pm\Gamma _{\pm }e^{i(k_{\Delta}d-\phi_{k_{\Delta}})}e^{2ik_{\Delta}m_{1}}/2}{i(\Delta
_{k}-\Delta _{\pm })-\Gamma _{\pm }/2},\label{AABB rpm}
\end{eqnarray}
with
\begin{subequations}
\begin{eqnarray}
\Delta _{\pm }&=&\Gamma_{e}\{\sin (k_{\Delta}d) \pm[1+\cos (k_{\Delta}d)]\sin(2k_{\Delta}d-\phi_{k_{\Delta}})\},\label{AABB Lamb shift pm} \\
\Gamma _{\pm }&=&2\Gamma_{e}[1+\cos (k_{\Delta}d)][1\pm \cos(2k_{\Delta}d-\phi_{k_{\Delta}})].\label{AABB individual decay pm}
\end{eqnarray}
\end{subequations}

We can further demonstrate that, under the condition $\Gamma _{\pm}\gg \Gamma _{\mp}$, the reflection spectra around $\Delta _{\mp}$ can be fitted by the Fano line shapes as described by Eq.~(\ref{Fano line shape}). According to Eq.~(\ref{AABB individual decay pm}), we find that when the dimerization parameters $\delta=0.5$ ($-0.5$) and the frequency detuning $\Delta$ is taken as $\Delta/J\in(1.14,1.24)\cup(1.24,1.34)\cup(1.84,1.90)\cup(1.90,1.94)$ [$(1,1.01)\cup(1.31,1.39)\cup(1.39,1.47)\cup(1.87,1.92)\cup(1.92,1.95)$], we have $\Gamma _{+}\gg\Gamma _{-}$, and then the reflection spectra exhibit the Fano line shapes around the minimal reflection, as shown by the red solid curve in Fig.~\ref{RAABBvsDeltakchgDelta}(c). When the dimerization parameters $\delta=0.5$ ($-0.5$) and $\Delta/J\in(1,1.02)\cup(1.52,1.58)\cup(1.58,1.71)\cup(1.99,2)$ [$(1.06,1.11)\cup(1.11,1.18)\cup(1.62,1.70)\cup(1.70,1.77)\cup(1.99,2)$], the width satisfies $\Gamma _{-}\gg\Gamma _{+}$. In this case, the reflection spectra exhibit the Fano line shapes around the minimal reflection, as shown by the purple dot-dashed curve in Fig.~\ref{RAABBvsDeltakchgDelta}(c) [the red solid curve in Fig.~\ref{RAABBvsDeltakchgDelta}(d)]. Here, the sign of $\delta$ affects the regions of the Fano line shapes, which is different from both the AAAA- and ABAB-coupling configurations.
%%%%%%%%%%%%%%%%%%%%%%%%%%%%%
\begin{figure}[tbp]
\center\includegraphics[width=0.48\textwidth]{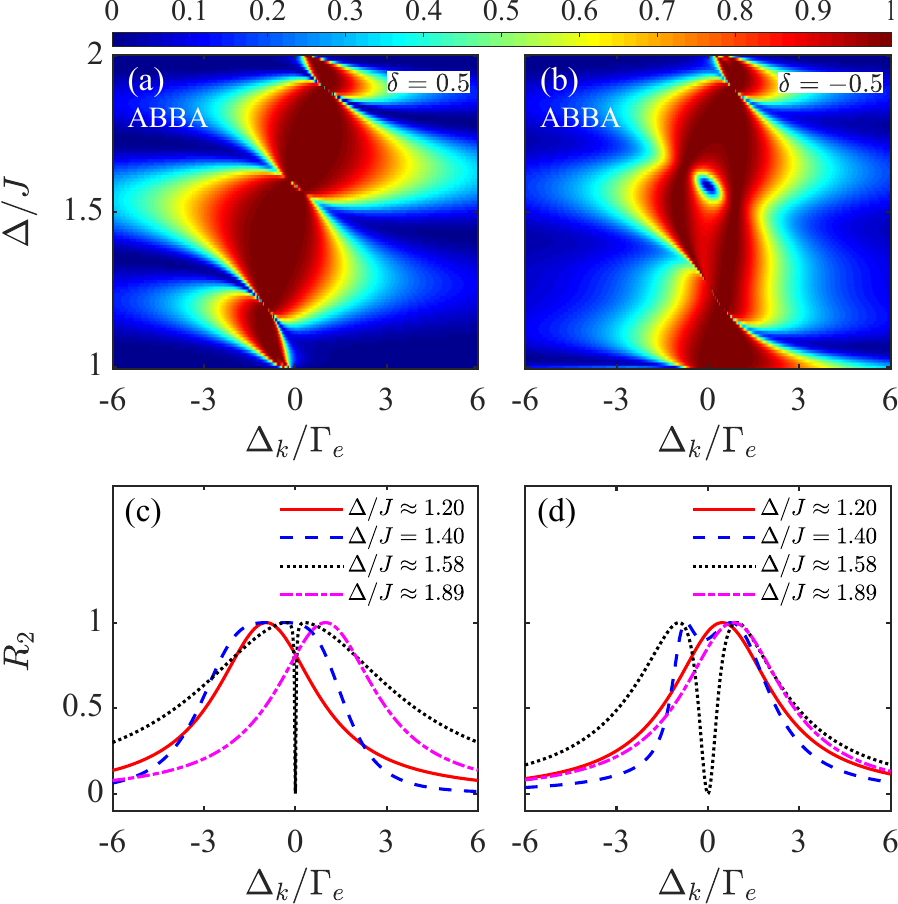}
\caption{Reflection coefficient $R_{2}$ for the ABBA-coupling configuration as a function of $\Delta _{k}$ and $\Delta$. The dimerization parameter is (a) $\delta=0.5$ and (b) $\delta=-0.5$. The profiles of panels (a) and (b) at several values of $\Delta$ are shown by the curves in (c) and (d). Other parameters used are $g=0.01J$ and $d=2$.}
\label{RABBAvsDeltakchgDelta}
\end{figure}
%%%%%%%%%%%%%%%%%%%%%%%%%%%%%

When $\delta=0.5$ ($-0.5$) and $\Delta\approx1.24J$ (the blue dashed curves), the reflection spectrum shows the Lorentzian line shape (the super-Gaussian line shape). Specifically,  for the Lorentzian line shape ($\delta=0.5$ and $\Delta\approx1.24J$), we have the characteristic parameters $\Delta_{\text{S}L,k_{\Delta}}=\Delta_{\text{A}L,k_{\Delta}}\approx0.99\Gamma_{e}$, $\Gamma_{\text{S},k_{\Delta}} \approx3.38\Gamma_{e}$, $\Gamma_{\text{A},k_{\Delta}}\approx0$, and $g_{\text{SA},k_{\Delta}}=\Gamma_{\text{SA},k_{\Delta}}=0$. In this case, the antisymmetric (symmetric) state is decoupled from (coupled to) the SSH waveguide. The reflection spectrum exhibits the Lorentzian line shape centered at $\Delta _{k}=0.99\Gamma _{e}$ with the linewidth of $3.38\Gamma _{e}$, as shown by the blue dashed curve in Fig.~\ref{RAABBvsDeltakchgDelta}(c). For the super-Gaussian line shape, the reflection coefficient around the reflection peaks $\Delta _{k}=\Delta _{L2,k_{\Delta}}=\Delta _{L1,k_{\Delta}}=\Delta _{L,k_{\Delta}}$ can be approximated as
\begin{eqnarray}
R_{2} &\approx& 1-\Delta'/(4g_{12,k_{\Delta}}^{4}+\Gamma _{12,k_{\Delta}}^{2}g_{12,k_{\Delta}}^{2}) \notag \\
&\approx & \text{exp}[-\Delta'/(4g_{12,k_{\Delta}}^{4}+\Gamma _{12,k_{\Delta}}^{2}g_{12,k_{\Delta}}^{2})],\label{super-Gaussian line shapes}
\end{eqnarray}
where $\Delta'=\Delta_{k}-\Delta _{L,k_{\Delta}}$. Equation~(\ref{super-Gaussian line shapes}) indicates that the reflection spectrum around the reflection peak exhibits a super-Gaussian line shape, as shown by the blue dashed curve in Fig.~\ref{RAABBvsDeltakchgDelta}(d). When $\delta=\pm0.5$ and $\Delta\approx1.58J$ (the black dotted curves), the characteristic quantities become $\Delta_{\text{S}L,k_{\Delta}}=\Delta_{\text{A}L,k_{\Delta}}=\Gamma_{\text{S},k_{\Delta}} =\Gamma_{\text{A},k_{\Delta}}=g_{\text{SA},k_{\Delta}}=\Gamma_{\text{SA},k_{\Delta}}=0$. Therefore, both the symmetric and antisymmetric states are decoupled from the waveguide and then the single photon is completely transmitted. This feature is consistent with the AAAA-coupling configuration. When $\delta=\pm0.5$ and $\Delta\approx1.70J$ (the purple dot-dashed curves), the reflection spectra show the asymmetric Fano ($\Gamma _{-}\gg\Gamma _{+}$) and Lorentzian line shape, respectively. For the Lorentzian line shape ($\delta=-0.5$ and $\Delta\approx1.70J$), we have the characteristic parameters $\Delta_{\text{S}L,k_{\Delta}}=\Delta_{\text{A}L,k_{\Delta}}\approx-0.49\Gamma_{e}$, $\Gamma_{\text{S},k_{\Delta}} \approx0$, $\Gamma_{\text{A},k_{\Delta}}\approx0.51\Gamma_{e}$, and $g_{\text{SA},k_{\Delta}}=\Gamma_{\text{SA},k_{\Delta}}=0$. In this case, the symmetric (antisymmetric) state is decoupled from (coupled to) the SSH waveguide. Therefore, the reflection spectrum exhibits the Lorentzian line shape centered at $\Delta _{k}=-0.49\Gamma _{e}$ with the linewidth of $0.51\Gamma _{e}$, as shown by the purple dot-dashed curve in Fig.~\ref{RAABBvsDeltakchgDelta}(d).

\subsubsection{The ABBA-coupling configuration}
For the ABBA-coupling configuration, the dimensionless parameters in Eqs.~(\ref{t2})--(\ref{exchange interaction}) are $\mu_{1(4)}=\nu_{2(3)}=1$ and $\nu_{1(4)}=\mu_{2(3)}=0$. In Figs.~\ref{RABBAvsDeltakchgDelta}(a) and \ref{RABBAvsDeltakchgDelta}(b), we plot the reflection coefficient as a function of $\Delta _{k}$ and $\Delta$. Different from the previous three coupling cases, it can be seen that the reflection spectra have two reflection peaks ($R_{2}=1$) at $\Delta_{k}=\Gamma _{e}\sin(k_{\Delta}d-\phi _{k_{\Delta}})$ and $\Delta_{k}=\Gamma _{e}\sin(k_{\Delta}d+\phi _{k_{\Delta}})$. Here, the locations of the two peaks depend on the sign of $\delta$. To show more details, we plot in Figs.~\ref{RABBAvsDeltakchgDelta}(c) and \ref{RABBAvsDeltakchgDelta}(d) the corresponding profiles of Figs.~\ref{RABBAvsDeltakchgDelta}(a) and \ref{RABBAvsDeltakchgDelta}(b) at various values of $\Delta$, respectively.

To analyze the characteristics of the reflection spectra more clearly, we substitute $\delta=\pm0.5$ and $\Delta\approx1.20J$ (the red solid curves) into Eqs.~(\ref{r2})--(\ref{exchange interaction}) and obtain the reflection rate
\begin{eqnarray}
R_{2}^{\text{ABBA}}=\frac{(\Gamma _{1,k_{\Delta}}+\Gamma _{2,k_{\Delta}})^{2}/4}{(\Delta_{k}-\Delta_{L,k_{\Delta}})^{2}+(\Gamma _{1,k_{\Delta}}+\Gamma _{2,k_{\Delta}})^{2}/4}.\label{ABBA Lor}
\end{eqnarray}
Here, the characteristic parameters become $\Delta _{L1,k_{\Delta}}=\Delta _{L2,k_{\Delta}}=\Delta _{L,k_{\Delta}}\approx-0.96\Gamma_{e}$ ($0.46\Gamma_{e}$), $\Gamma _{1,k_{\Delta}}+\Gamma _{2,k_{\Delta}}=4\Gamma_{e}$, $g_{12,k_{\Delta}}\approx0$, and $\Gamma _{1,k_{\Delta}}\Gamma _{2,k_{\Delta}}-\Gamma_{12,k_{\Delta}}^{2}\approx0$ for $\delta=\pm0.5$. Therefore, the reflection spectrum exhibits the Lorentzian line shapes centered at $\Delta _{k}=-0.96\Gamma_{e}$ ($0.46\Gamma_{e}$) with the linewidth of $4\Gamma _{e}$, as shown by the red solid curves in Figs.~\ref{RABBAvsDeltakchgDelta}(c) and \ref{RABBAvsDeltakchgDelta}(d), respectively. When $\delta=0.5$ ($-0.5$) and $\Delta\approx1.89J$, the characteristic quantities become $\Delta _{L1,k_{\Delta}}=\Delta _{L2,k_{\Delta}}=\Delta _{L,k_{\Delta}}\approx0.98\Gamma_{e}$ ($0.83\Gamma_{e}$), $\Gamma _{1,k_{\Delta}}+\Gamma _{2,k_{\Delta}}=4\Gamma_{e}$, $g_{12,k_{\Delta}}\approx0$, and $\Gamma _{1,k_{\Delta}}\Gamma _{2,k_{\Delta}}-\Gamma_{12,k_{\Delta}}^{2}\approx0$, which satisfy Eq.~(\ref{ABBA Lor}), as shown by the purple dot-dashed curve in Fig.~\ref{RABBAvsDeltakchgDelta}(c) [Fig.~\ref{RABBAvsDeltakchgDelta}(d)]. In other parameters, the reflection spectra have two completely reflection peaks, such as $\Delta=1.40J$ (the blue dashed curves). In particular, when $\delta=0.5$ $(-0.5)$ and $\Delta\approx1.58J$, we have $\Delta _{\text{S}L,k_{\Delta}}\approx\Delta _{\text{A}L,k_{\Delta}}\approx0$, $\Gamma _{\text{S},k_{\Delta}}\approx0$, $\Gamma _{\text{A},k_{\Delta}}\approx7.79\Gamma_{e}$ $(5.26\Gamma_{e})$, $g_{\text{SA},k_{\Delta}}\approx-0.32\Gamma_{e}$ $(-0.95\Gamma_{e})$, and $\Gamma _{\text{SA},k_{\Delta}}\approx0$. Here, the exchange interaction strength $g_{\text{SA},k_{\Delta}}$ and the individual decay rate of the antisymmetric state $\Gamma _{\text{A},k_{\Delta}}$ satisfy the condition $|g_{\text{SA},k_{\Delta}}|<\Gamma _{\text{A},k_{\Delta}}/4$~\cite{Feng2021,Abi-Salloum2010,Anisimov2011}. In this case, the system exhibits an electromagnetically induced transparency (EIT)-like spectrum. Thus we find that the reflection spectrum has a minimal value ($R_{2}\approx0$) at $\Delta _{k}\approx0$ and that the reflection spectrum has two symmetric complete-reflection peaks ($R_{2}=1$) at $\Delta _{k}\approx\pm0.32\Gamma_{e}$ $(\pm0.95\Gamma_{e})$, as shown by the black dotted curve in Fig.~\ref{RABBAvsDeltakchgDelta}(c) [Fig.~\ref{RABBAvsDeltakchgDelta}(d)].
%%%%%%%%%%%%%%%%%%%%%%%%%%%%%
\begin{figure}[tbp]
\center\includegraphics[width=0.48\textwidth]{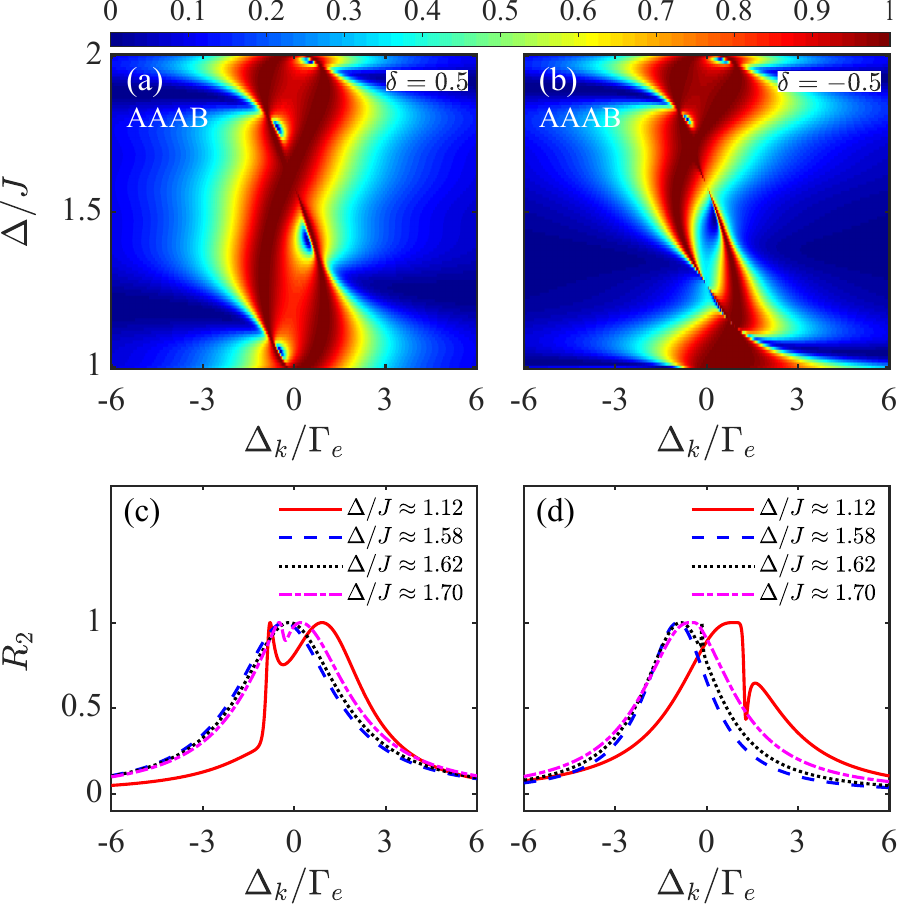}
\caption{Reflection coefficient $R_{2}$ for the AAAB-coupling configuration as a function of $\Delta _{k}$ and $\Delta$. The dimerization parameter is (a) $\delta=0.5$ and (b) $\delta=-0.5$. The profiles of panels (a) and (b) at several values of $\Delta$ are shown by the curves in (c) and (d). Other parameters used are $g=0.01J$ and $d=2$.}
\label{RAAABvsDeltakchgDelta}
\end{figure}
%%%%%%%%%%%%%%%%%%%%%%%%%%%%%

\subsubsection{The AAAB-coupling configuration}
For the AAAB-coupling configuration, the dimensionless parameters in Eqs.~(\ref{t2})--(\ref{exchange interaction}) are given by $\mu_{1-3}=\nu_{4}=1$ and $\nu_{1-3}=\mu_{4}=0$. In Figs.~\ref{RAAABvsDeltakchgDelta}(a) and \ref{RAAABvsDeltakchgDelta}(b), we show the reflection coefficient as a function of $\Delta _{k}$ and $\Delta$. We find that the reflection spectra have two reflection peaks ($R_{2}=1$) at $\Delta_{k}=\Delta _{L1,k_{\Delta}}=\Gamma _{e}\sin(k_{\Delta}d)$ (the first reflection peak) and $\Delta_{k}=\Delta _{L2,k_{\Delta}}=\Gamma _{e}\sin(k_{\Delta}d-\phi _{k_{\Delta}})$ (the second reflection peak). Here, the second Lamb shift $\Delta _{L2,k_{\Delta}}$ depends on $\phi _{k_{\Delta}}$ and hence the position of the second reflection peak is influenced by the sign of $\delta$. To show the details clearly, in Figs.~\ref{RAAABvsDeltakchgDelta}(c) and \ref{RAAABvsDeltakchgDelta}(d) we plot the profiles of Figs.~\ref{RAAABvsDeltakchgDelta}(a) and \ref{RAAABvsDeltakchgDelta}(b) at different frequency detunings in the region $\Delta \in(J,2J)$, respectively.

When $\delta=\pm0.5$ and $\Delta\approx1.58J$ (the blue dashed curves), the reflection spectra exhibit the Lorentzian line shapes. Similarly, we substitute $\delta=\pm0.5$ and $\Delta\approx1.58J$ into Eqs.~(\ref{r2})--(\ref{exchange interaction}) and obtain the reflection coefficient as
\begin{eqnarray}
R_{2}^{\text{AAAB}}=\frac{\Gamma _{2,k_{\Delta}}^{2}/4}{(\Delta_{k}-\Delta_{L2,k_{\Delta}})^{2}+\Gamma _{2,k_{\Delta}}^{2}/4}.\label{AAAB Lor}
\end{eqnarray}
Here, we have $\Delta _{L1,k_{\Delta}}=0$, $\Delta _{L2,k_{\Delta}}\approx-0.32\Gamma_{e}$ ($-0.95\Gamma_{e}$), $\Gamma _{1,k_{\Delta}}=0$, $\Gamma _{2,k_{\Delta}}\approx3.9\Gamma_{e}$ ($2.6\Gamma_{e}$), $g_{12,k_{\Delta}}\approx0$, and $\Gamma_{12,k_{\Delta}}\approx0$ for $\delta=0.5$ ($-0.5$). Therefore, the first giant atom is decoupled from the waveguide, while the second giant atom is coupled to the waveguide. In this case, the reflection spectrum exhibits the Lorentzian line shape centered at $\Delta _{k}=-0.32\Gamma_{e}$ ($-0.95\Gamma_{e}$) with the linewidth of $3.9\Gamma_{e}$ ($2.6\Gamma_{e}$), as shown by the blue dashed curve in Fig.~\ref{RAAABvsDeltakchgDelta}(c) [Fig.~\ref{RAAABvsDeltakchgDelta}(d)]. In particular, when $\delta=\pm0.5$, $\Delta\approx1.12J$ (the red solid curves), $1.62J$ (the black dotted curves), and $1.70J$ (the purple dot-dashed curves), the number of the reflection peak depends on the sign of the dimerization parameter $\delta$. Specifically, when $\Delta\approx1.12J$ (the red solid curves) and $1.70J$ (the purple dot-dashed curves); the reflection spectra at $\delta>0$ have two reflection peaks, while the reflection spectra at $\delta<0$ have one reflection peak and vice versa at $\Delta\approx1.62J$ (the black dotted curves). This is because the two reflection peaks at $\Delta _{L1,k_{\Delta}}$ and $\Delta _{L2,k_{\Delta}}$ coincide when $\delta=-0.5$ and $\Delta\approx1.12J$ or $1.70J$ (when $\delta=0.5$ and $\Delta\approx1.62J$), i.e., $\Delta _{k}=\Gamma _{e}\sin(k_{\Delta}d)=\Gamma _{e}\sin(k_{\Delta}d-\phi _{k_{\Delta}})$. In this case, the reflection spectra have one reflection peak. These analyses indicate that only the position of the second peak depends on the sign of $\delta$. When the sign of $\delta$ changes, the position of the second peak changes accordingly.
%%%%%%%%%%%%%%%%%%%%%%%%%%%%%
\begin{figure}[tbp]
\center\includegraphics[width=0.48\textwidth]{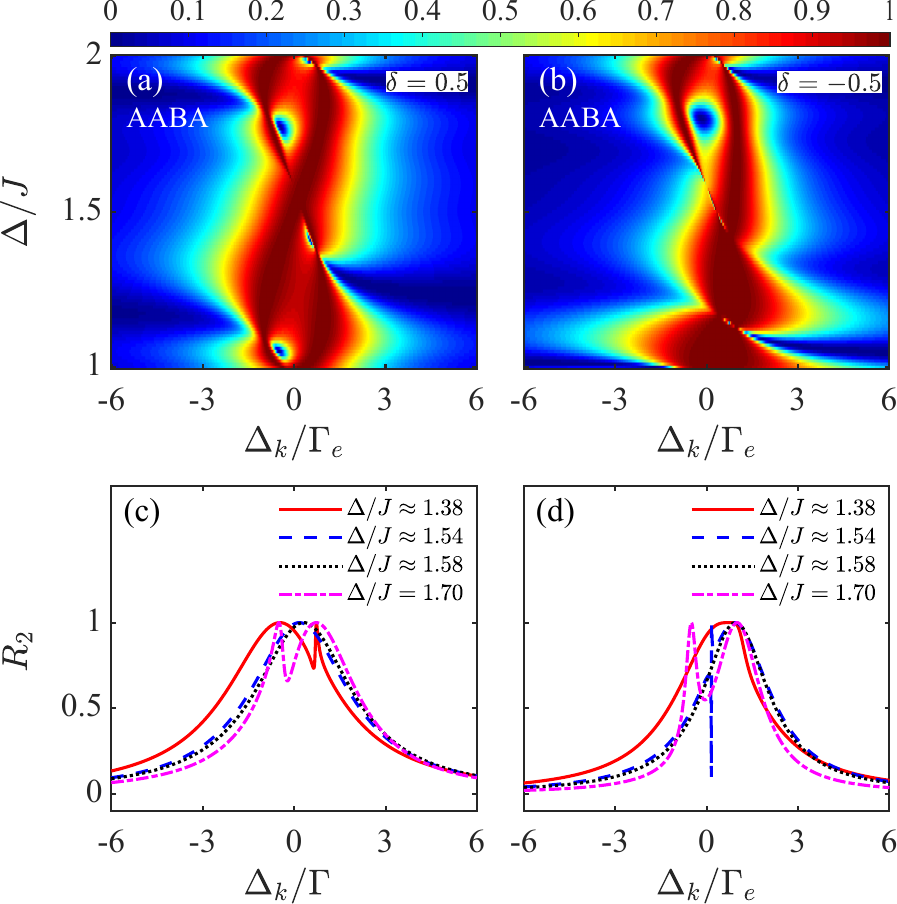}
\caption{Reflection coefficient $R_{2}$ for the AABA-coupling configuration as a function of $\Delta _{k}$ and $\Delta$. The dimerization parameter is (a) $\delta=0.5$ and (b) $\delta=-0.5$. The profiles of panels (a) and (b) at several values of $\Delta$ are shown by the curves in (c) and (d). Other parameters used are $g=0.01J$ and $d=2$.}
\label{RAABAvsDeltakchgDelta}
\end{figure}
%%%%%%%%%%%%%%%%%%%%%%%%%%%%%

\subsubsection{The AABA-coupling configuration}
For the AABA-coupling configuration, the dimensionless parameters in Eqs.~(\ref{t2})--(\ref{exchange interaction}) are $\mu_{1,2,4}=\nu_{3}=1$ and $\nu_{1,2,4}=\mu_{3}=0$. Figures~\ref{RAABAvsDeltakchgDelta}(a) and \ref{RAABAvsDeltakchgDelta}(b) show the reflection coefficient as a function of $\Delta _{k}$ and $\Delta$. It can be found that the reflection spectra have two peaks at $\Delta_{k}=\Gamma _{e}\sin(k_{\Delta}d)$ (the first reflection peak) and $\Delta_{k}=\Gamma _{e}\sin(k_{\Delta}d+\phi _{k_{\Delta}})$ (the second reflection peak), respectively. Therefore, only the position of the second peak depends on the sign of $\delta$, which is similar to the AAAB-coupling configuration. In Figs.~\ref{RAABAvsDeltakchgDelta}(c) and \ref{RAABAvsDeltakchgDelta}(d) we show the profiles of Figs.~\ref{RAABAvsDeltakchgDelta}(a) and \ref{RAABAvsDeltakchgDelta}(b) at different frequency detunings in the region $\Delta \in(J,2J)$, respectively.

When $\delta=\pm0.5$ and $\Delta\approx1.58J$ (the black dotted curves), the reflection spectra exhibit the Lorentzian line shapes, which are consistent with the AAAB-coupling configuration. According to Eqs.~(\ref{r2})--(\ref{exchange interaction}), we can obtain the reflection coefficient $R_{2}^{\text{AABA}}$, which satisfies Eq.~(\ref{AAAB Lor}). Here, the characteristic quantities become $\Delta _{L1,k_{\Delta}}=0$, $\Delta _{L2,k_{\Delta}}\approx0.32\Gamma_{e}$ ($0.95\Gamma_{e}$), $\Gamma _{1,k_{\Delta}}=0$, $\Gamma _{2,k_{\Delta}}\approx3.9\Gamma_{e}$ ($2.6\Gamma_{e}$), $g_{12,k_{\Delta}}\approx0$, and $\Gamma_{12,k_{\Delta}}\approx0$ for $\delta=0.5$ ($-0.5$). Therefore, the first giant atom is decoupled from the waveguide, while the second giant atom is coupled to the waveguide. In this case, the reflection spectrum exhibits the Lorentzian line shape centered at $\Delta _{k}=0.32\Gamma_{e}$ ($0.95\Gamma_{e}$) with the linewidth of $3.9\Gamma_{e}$ ($2.6\Gamma_{e}$), as shown by the black dotted curve in Fig.~\ref{RAABAvsDeltakchgDelta}(c) [Fig.~\ref{RAABAvsDeltakchgDelta}(d)]. From the above analyses, it can be seen that, when $\delta=0.5$ ($-0.5$) and $\Delta\approx1.58J$, the reflection spectra of AAAB- and AABA-coupling configurations are symmetric with respect to $\Delta_{k}=0$, as shown by the blue dashed curve in Fig.~\ref{RAAABvsDeltakchgDelta}(c) [Fig.~\ref{RAAABvsDeltakchgDelta}(d)] and the black dotted curve in Fig.~\ref{RAABAvsDeltakchgDelta}(c) [Fig.~\ref{RAABAvsDeltakchgDelta}(d)]. For other values of $\Delta$, such as $\Delta=1.70J$, the reflection spectra have two reflection peaks [see the purple dot-dashed curves in Figs.~\ref{RAABAvsDeltakchgDelta}(c) and \ref{RAABAvsDeltakchgDelta}(d)]. In particular, when $\delta=\pm0.5$, $\Delta\approx1.38J$ (the red solid curves), and $1.54J$ (the blue dashed curves), the number of the reflection peaks depends on the sign of $\delta$. Specifically, when $\Delta\approx1.38J$ (the red solid curves), the reflection spectrum at $\delta>0$ has two reflection peaks, while the reflection spectrum at $\delta<0$ has one reflection peak and vice versa at $\Delta\approx1.54J$ (the blue dashed curves). When $\delta=-0.5$ and $\Delta\approx1.38J$ ($\delta=0.5$ and $\Delta\approx1.54J$), we have the relation $\Delta _{k}=\Gamma _{e}\sin(k_{\Delta}d)=\Gamma _{e}\sin(k_{\Delta}d+\phi _{k_{\Delta}})$. Thus the reflection spectra have one reflection peak, which is similar to the AAAB-coupling configuration.

\section{Conclusion}\label{Conclusion}
In conclusion, we have studied the single-photon scattering in an SSH waveguide coupled to either one or two two-level giant atoms. In both cases, we have derived the unified analytical expressions of the single-photon scattering amplitudes for different coupling configurations. We have found that the single-photon scattering is determined by the coupling configurations, coupling distances, atomic resonance frequency, and dimerization parameter. By analyzing the analytical expressions of the scattering amplitudes, we have found that, when all (any two or more) legs of the giant atoms are coupled to the sublattices of the same (different) type, the scattering behavior is independent of (depends on) the sign of dimerization parameter.

For photons scattered by a single two-level giant atom, we have considered four different coupling configurations according to the coupling-point distributions. In particular, we have shown that the scattering spectra can show different line shapes by adjusting jointly the quantum interference effect and topological effect. In the case of two giant atoms, we have considered the single-photon scattering for sixteen coupling configurations. By comparing the Lamb shift, exchanging interaction, individual decay, and collective decay, we have further simplified these sixteen coupling configurations to six coupling cases. In particular, the single-photon scattering spectra can exhibit Lorentzian, super-Gaussian, EIT-like, and asymmetric Fano line shapes for some coupling configurations. The parameter ranges for the appearance of these line shapes have been discussed in detail, which can provide a valuable reference for the experimental study of giant-atom topological-waveguide-QED systems. These results show that the topological-waveguide-QED systems with giant atoms can not only  exhibit a quantum switch for coherent single-photon transport, but also serve as a platform for probing the topological characteristics of these waveguide systems. This work will pave the way for the study of controllable single-photon devices in giant-atom topological-waveguide-QED systems.

\begin{acknowledgments}
J.-Q.L. was supported in part by the National Natural Science Foundation of China (Grants No.~12175061, No.~12247105, No.~11935006, and No.~12421005), the National Key Research and Development Program of China (Grant No.~2024YFE0102400), and the Hunan Provincial Major Sci-Tech Program (Grant No.~2023ZJ1010). X.-L.Y. was supported in part by the Hunan Provincial Postgraduate Research and Innovation project (Grant No.~CX20230463).
\end{acknowledgments}


\begin{thebibliography}{99}
\bibitem{conditions} Due to the two sets of sublattices appearing alternately, the boundary conditions of the topological waveguide are different from those for the conventional waveguide. The boundary conditions vary with the coupling configuration. Here, we divide the probability amplitude into three parts: outside the coupling points, inside the coupling points, and at the coupling points.
\end{thebibliography}
\end{document}